\documentclass[american,twoside]{article}
\usepackage{a4wide}
\usepackage[T1]{fontenc}
\usepackage[latin1]{inputenc}
\usepackage{latexsym}
\usepackage{bbm}
\usepackage{babel}
\usepackage{fourier}
\usepackage{epsfig}
\usepackage{enumerate}
\usepackage{color}
\usepackage[active]{srcltx}
\usepackage[colorlinks=true]{hyperref}
\usepackage{fancyhdr}
\usepackage{mathrsfs}
\usepackage{amsmath,amsfonts,amssymb,amsthm}
\usepackage{fancyhdr}
\usepackage{graphicx}
\usepackage{bbm}
\usepackage{cancel}
\usepackage{nth}

\def\bel{\begin{lemma}}
\def\eel{\end{lemma}}
\def\bet{\begin{theoreme}}
\def\eet{\end{theoreme}}
\def\bed{\begin{definition}}
\def\eed{\end{definition}}
\def \beq{\begin{equation}}
\def \eeq{\end{equation}}

\def \beq{\begin{equation}}
\def \eeq{\end{equation}}

\def\supp{\mathop{\rm supp} \nolimits} 

\def\and {{\rm \; and \;}}

\def\loc{{\rm loc}}

\renewcommand{\Im}{{\rm Im}\,}
\renewcommand{\Re}{{\rm Re}\,}
\newcommand{\Ker}{{\rm Ker}}
\newcommand{\Ran}{{\rm Ran}}

\newcommand{\R}{{\mathbb R}}

\def\cS{{\cal S}}
\def\cR{{\cal R}}
\def\cH{{\cal H}}
\def\cO{{\cal O}}
\def\cB{{\cal B}}
\def\cT{{\cal T}}

\def\fh{{\mathfrak h}}
\def\fH{{\mathfrak H}}

\def\fS{{\mathfrak S}}

\def\rr{{\mathbb R}}
\def\cc{{\mathbb C}}
\def\one{{\mathbbm 1}}
\def\d{{\rm d}}
\def\e{{\rm e}}
\def\i{{\rm i}}
\def\tr{{\rm tr}}

\def\ds{\displaystyle}
\def\CAR{{\rm CAR}}
\def\CARv{{\rm CAR}_\vartheta}
\def\bbeta{{\boldsymbol{\beta}}}
\def\bmu{{\boldsymbol{\mu}}}

\def\b0{{\boldsymbol{0}}}
\def\st{{\langle\,\cdot\,\rangle}}

\def\jth{{$j^\mathrm{th}\,$}}

\newtheorem{theorem}{Theorem}[section]
\newtheorem{definition}[theorem]{Definition}
\newtheorem{proposition}[theorem]{Proposition}

\newtheorem{lemma}[theorem]{Lemma}

\theoremstyle{definition}
\newtheorem{remark}[theorem]{Remark}

\numberwithin{equation}{section}

\begin{document}
\def\today{}
\title{\bf {A mathematical account of the NEGF formalism}}
\author{\sc H.D. Cornean$^{1}$, V. Moldoveanu$^{2}$, C.-A. Pillet$^3$
\\ \\ \\
$^1$Department of Mathematical Sciences\\ 
Aalborg University\\
Fredrik Bajers Vej 7G, 9220 Aalborg, Denmark
\\ \\
$^2$National Institute of Materials Physics\\
P.O. Box MG-7  Bucharest-Magurele, Romania
\\ \\
$^3$
Aix Marseille Univ, Universit\'e de Toulon, CNRS, CPT, Marseille, France  
}
\maketitle
\thispagestyle{empty}
\begin{quote}
{\bf Abstract.} The main goal of this paper is to put on solid mathematical
grounds the so-called  Non-Equilibrium Green's Function (NEGF) transport
formalism for open systems.  In particular, we derive the Jauho-Meir-Wingreen 
formula for the time-dependent current through an interacting sample coupled to
non-interacting leads. Our proof is non-perturbative and uses neither
complex-time Keldysh contours, nor Langreth rules of `analytic continuation'. We
also discuss other technical identities (Langreth, Keldysh) involving various
many body Green's functions. Finally, we study the Dyson equation for the
advanced/retarded interacting Green's function and we rigorously construct its
(irreducible) self-energy, using the theory of Volterra operators.
\end{quote}

\maketitle
\thispagestyle{empty}

\section{Introduction}
The  computation of Green's functions (GFs) associated to interacting many-body
quantum systems is one of the most fruitful and challenging problems in
theoretical condensed matter physics. Among many applications of Green's functions
techniques let us mention the linear response to an adiabatically switched
perturbation (e.g.\;an external electric field or electron--electron
interactions) which is embodied in a single or two-particle GF. Also, the
equilibrium properties of an interacting system can be studied in the framework
of the finite-temperature (Matsubara) formalism. The great advantage of GFs is
that they can be obtained from perturbative approximations which satisfactorily
capture plenty of physical phenomena. In some sense they are much simpler
objects than $N$-particle wavefunctions which are difficult to compute even with
the currently available hardware. It was already clear in the early '60s that
non-equilibrium regimes cannot be described within ground-state perturbation
theory based on the Gell-Mann and Low Theorem~\cite{GML} as the system does not
return to the same initial state after the external driving is turned off.

In spite of this difficulty considerable efforts were spent to adapt the very
appealing and successful diagrammatic expansion of the ground-state GF (see,
e.g., \cite{FW}) to the non-equilibrium setting and in particular to transport
calculations for open quantum systems. The resulting theory of non-equilibrium
Green's functions (NEGFs) nowadays surpasses in predictive power quantum kinetic
(Boltzmann) equations\footnote{The Kadanoff-Baym equations lead to the Boltzmann
equation under appropriate assumptions, see the review~\cite{Da}.},
Kubo-Greenwood formulas~\cite{Ku} and the Landauer-B\"{u}ttiker approach to
coherent transport (see, e.g., \cite{Im}). Although the NEGF formalism was
proposed and developed in the '60s by several authors~\cite{Sc,KB,Ke,Fu,Cr} to
the best of our knowledge a review has not been available until '84~\cite{Da}.
Also, the first application of the Keldysh formalism to a transport problem has
been presented in a series of papers by Caroli et al.~\cite{Ca1, Ca2,Co,Ca3}. It
is interesting to note that in their first paper the authors introduced the
partitioning transport setting which is nowadays standard textbook material (the
setting will be described in the next section). The real breakthrough of the
non-equilibrium GFs method to transport problems is due to Meir and
Wingreen~\cite{MW}. They proposed a closed formula for the steady-state current
across an interacting region which was then extended to the transient
regime~\cite{JWM}.

Given the fact that in the non-interacting case the single-particle GF reduces
to the resolvent of the full Hamiltonian which can either be computed in
perturbation theory or related to scattering theory, it is not surprising that
some aspects of Green's functions methods have been also addressed from the
mathematical point of view. Two such examples are the Green-Kubo formulation of
linear response theory and the Landauer-B\"uttiker approach to coherent
transport in mesoscopic systems. Both methods were rigorously established for
independent electron (i.e., quasi-free) models with discrete and continuous 
geometries~\cite{AH,AP,AJPP2,BP,CJM,CDNP,N}.

In the interacting case, and more generally for non quasi-free dynamics,
mathematical constructions of current carrying steady states have been
obtained within the perturbative approach~\cite{FMU,JP1,AJPP1,MMS}. 
The linear response theory of these nonequilibrium steady states (NESS)
is also well understood at the mathematical level~\cite{JOP1,JOP2,JOP3}.
Using the scattering theoretical approach developed in these works, we have
recently obtained steady-states limits of the two-point  GFs of an open mesoscopic
sample~\cite{CMP1,CMP2,CM} for various, partitioning and non-partitioning 
protocols. We have also presented an alternative perturbative formula for the
steady-state current and established the independence of the steady-state
quantities on the initial state of the sample.

As a natural continuation of these results the present work aims at providing
the first rigorous account of the Keldysh-Green's functions machinery which is
rather  inaccessible to the mathematical community in its original formulation.
Indeed, for a beginner or a mathematically oriented reader the first  unfamiliar
technical object of the NEGF formalism is quite its starting point: the famous
Keldysh contour introduced along with its associated countour-ordering operator
which replaces the time-ordering operator of the equilibrium theory. In the
simplest version of the theory (i.e., if the initial correlations are neglected)
the contour runs from some initial time $t_0$ where the system is still in an
{\it equilibrium} state to some later time $t$ and then back to $t_0$. This
unusual choice becomes even stranger when one adds a misleading picture in which
the contour is slightly extended to the complex plane. If the initial state is
correlated the Keldysh contour has an additional complex (Matsubara) `hook'
($t\in [t_0-\i\beta,t_0]$).  Let us emphasize that this construction plays a 
crucial role in transport calculations where one uses formal identities for GFs 
whose time variables are seen on different `branches' of the Keldysh contour.

If one carefully follows the development of the many-body perturbative
approaches, {\it the only} `raison d'\^{e}tre' of the Keldysh contour is
quickly unraveled: it provides a systematic and convenient way to order time
arguments of complicated products of interaction picture operators on the
chronological (from $t_0$ to $t$) and anti-chronological (from $t$ to $t_0$) time
branches. These products appear naturally in the statistical average of a given
observable in the non-equilibrium regime. A nice discussion on this point can be
found in the review by Danielewicz~\cite{Da}. Moreover, a compact form amenable
to a diagrammatic analysis via the Wick theorem is achieved by introducing
contour-ordered GFs, the latter being nothing but Keldysh GFs. To sum up, the
Keldysh contour helps one to reveal the formal analogy between the
equilibrium many-body perturbation theory and the non-equilibrium one.
Nevertheless, it is rather unfortunate that in order to study the
non-equilibrium, one has to be familiar to the ground-state perturbation theory,
Matsubara Green's functions, real-time GFs and so 
on~\cite{MSSL,Ne,ND,NDG,FVA}.  According to a recent
point of view~\cite{SvL}, the non-equilibrium many-body perturbation formalism is
the first one to be learned, the other ones being derived from it as simplified
versions. 

In our paper we establish a `contour-free' viewpoint of the NEGF formalism. A
certain advantage of this approach is that a mathematically oriented reader can
get to the JMW formula without an a priori knowledge of many-body theory or
Feynman diagrams. On the other hand, it should not come as a surprise that the
non-equilibrium transport can be formulated without contour-ordered quantities
given the fact that the seminal work of Kadanoff and Baym~\cite{KB} is written
down only in terms of retarded/advanced and correlation GFs.

The paper is organized as follows:

\begin{itemize}
\item After this Introduction we continue with a description of the setting in
Section~\ref{sectiunea2}. We are only dealing with discrete, partitioned
systems; the partition-free case~\cite{SA} will be treated elsewhere. Note that
we work right from the beginning at the thermodynamic limit, the only important
properties of the reference state being the KMS property and
gauge-invariance (i.e. it `commutes' with the number operator).

\item In Section~\ref{mainresults} we introduce the NEGFs and we formulate our
main results. Sections~\ref{prooofs} and~\ref{daison} contain all the  proofs.
Section~\ref{concluzii} concludes the paper and lists some open problems. In the
Appendix we prove a positivity lemma related to the dissipative properties of
the retarded interacting GF.

\item While in this work we shall use standard quantities (lesser,
retarded/advanced GFs) and tools (Wick theorem, Dyson equation) from many-body
perturbation theory, our proofs {\it do not} require two essential ingredients
of the Keldysh formalism: the contour-ordering operator and the so called
`analytic continuation' Langreth rules~\cite{La,HJ}. Although a significant
number of papers in the physical literature confirm the usefulness of these
methods in specific calculations based on diagrammatic expansion, our approach
shows that they are not mandatory for the study of non-equilibrium transport. In
fact we are able to provide in Theorem~\ref{jauho} the first rigorous proof of
the probably most famous by-product of the NEGF formalism, namely the
Jauho-Meir-Wingreen (JMW)~\cite{JWM} formula for the time-dependent current
through an interacting quantum dot without using contour-ordered GFs and
Langreth rules; the only technical ingredients we need are the well known KMS
condition~\eqref{multiKMS} and the Duhamel identity~\eqref{Duhamel0}. Along the
way we also present `contourless' derivations of the Langreth rules and of the
Keldysh equation for the lesser Green's function.

\item The JMW formula only involves lesser and retarded interacting GFs
restricted to the small sample. When the restriction to the small sample of the
initial state is the vacuum, we show in Proposition~\ref{Dkeldysh} how to
express lesser GFs in terms of non-interacting advanced/retarded GFs and a 
lesser self-energy, which in principle can be computed in all orders of the
interaction.

\item Our final result, Proposition~\ref{Dyson}, is a rigorous formulation of a
Dyson equation for the interacting advanced/retarded GFs, using the theory of
Volterra operators. In the proof of Lemma~\ref{lemaaprili} we also describe how
one can compute the self-energy in any order of the interaction and we
explicitly identify  its leading terms.
\end{itemize}

\medskip
\noindent \textbf{Acknowledgments}. The idea of a rigorous mathematical approach
to the NEGF formalism appeared more than ten years ago, catalyzed by lively
discussions with, among others, N.~Angelescu, J.~Derezi{\'n}ski, P.~Duclos, 
P.~Gartner, V.~Jak\v si\'c,  G.~Nenciu, G.~Stefanucci, and V.~Zagrebnov. Financial
support by Grant 4181-00042 of the Danish Council for Independent Research $|$
Natural Sciences is gratefully acknowledged. The work of CAP was partly funded by 
Excellence Initiative of Aix-Marseille University-A*MIDEX, a French 
``Investissements d'Avenir'' program. VM acknowledges financial support by the 
CNCS-UEFISCDI grant PN-III-P4-ID-PCE-2016-0221. 

\section{The model}
\label{sectiunea2}

In order to avoid technicalities which would only obscure the exposition we
restrict ourselves to a simple model of a discrete sample $\cS$ coupled to a
finite collection of reservoirs $\cR$ of spinless electrons within
the partitioning scenario used in our previous work~\cite{CMP2}. In this
section, we briefly recall this setup and refer the reader to~\cite{CMP2} for a
detailed discussion and results pertaining to the existence of steady states in
this model.

\subsection{The one-particle setup}
Our main object of interest is a Fermi gas on a discrete structure 
$\cS+\cR$ (e.g., an electronic system in the tight-binding approximation). 
There, $\cS$ is a finite set describing a confined sample and 
$\cR=\cR_1+\cdots+\cR_m$ is a collection of infinitely extended reservoirs (or 
leads) which feed the sample $\cS$.

The one-particle Hilbert space of the compound system is
$$
\fh=\fh_\cS\oplus\fh_\cR,\qquad \fh_\cR=\oplus_{j=1}^m\fh_j,
$$
where $\fh_\cS=\ell^2(\cS)$ and $\fh_j$ is the Hilbert space of the \jth
reservoir. Let $h_\cS$, a self-adjoint operator on
$\fh_\cS$, be the one-particle Hamiltonian of the isolated sample. Denote by 
$h_j$ the Hamiltonian of the \jth reservoir. The one-particle Hamiltonian of the 
decoupled system is
$$
h_{\rm D}=h_\cS\oplus h_\cR,\qquad h_\cR=\oplus_{j=1}^mh_j.
$$
The coupling of the sample to the reservoirs is achieved by the tunneling 
Hamiltonian
$$
h_{\rm T}=\sum_{j=1}^m d_j\left(|\psi_j\rangle\langle\phi_j|
+|\phi_j\rangle\langle\psi_j|\right),
$$
where $\psi_j\in\fh_j$ and $\phi_j\in\fh_\cS$ are unit vectors and 
$d_j\in\rr$ a coupling constant. The one-particle Hamiltonian of the fully
coupled system is
$$
h=h_{\rm D}+h_{\rm T}.
$$
In the following, we will denote by $1_{j/\cS/\cR}$ the orthogonal projection 
acting on the one-particle Hilbert space $\fh$ with range $\fh_{j/\cS/\cR}$.

\subsection{The many-body setup}
We shall now describe the Fermi gas associated with the one-particle model
introduced previously and extend this model by adding many-body interactions
between the particles in the sample $\cS$. In order to fix our notation and make
contact with the one used in the physics literature let us recall some basic 
facts. We refer to~\cite{BR2} for details on the algebraic framework of quantum
statistical mechanics that we use here.

$\Gamma_-(\fh)$ denotes the fermionic Fock space over $\fh$ and
$\Gamma_-^{(n)}(\fh)=\fh^{\wedge n}$, the $n$-fold completely antisymmetric
tensor power of $\fh$, is the $n$-particle sector of $\Gamma_-(\fh)$. For
$f\in\fh$, let $a(f)/a^\ast(f)$ be the annihilation/creation operator on
$\Gamma_-(\fh)$. In the following $a^\#$ stands for either $a$ or $a^\ast$. The
map $f\mapsto a^\ast(f)$ is linear while $f\mapsto a(f)$ is anti-linear, both
maps being continuous, $\|a^\#(f)\|=\|f\|$. The underlying algebraic structure
is characterized by the canonical anti-commutation relations (CAR for short)
\beq
\{a(f),a^\ast(g)\}=\langle f|g\rangle\one,\qquad\{a(f),a(g)\}=0,
\label{CAR}
\eeq
and we denote by $\CAR(\fh)$ the $C^\ast$-algebra generated by 
$\{a^\#(f)\,|\,f\in\fh\}$, i.e., the
norm closure of the set of polynomials in the operators $a^\#(f)$.
Note that if $\mathfrak{g}\subset\fh$ is a subspace, then we can identify 
$\CAR(\mathfrak{g})$
with a subalgebra of $\CAR(\fh)$.

The second quantization of a unitary operator $u$ on $\fh$ is the unitary 
$\Gamma(u)$ on $\Gamma_-(\fh)$ acting as $u\otimes u\otimes\cdots\otimes u$ on 
$\Gamma_-^{(n)}(\fh)$.
The second quantization of a self-adjoint operator $q$ on $\fh$ is the 
self-adjoint generator $\d\Gamma(q)$ of the strongly continuous unitary group 
$\Gamma(\e^{\i tq})$, i.e., $\Gamma(\e^{\i tq})=\e^{\i t\d\Gamma(q)}$. If 
$\{f_\iota\}_{\iota\in I}$ is an orthonormal basis of $\fh$ and $q$ a bounded 
self-adjoint operator, then
$$
\d\Gamma(q)=\sum_{\iota,\iota'\in I}\langle f_\iota|q| f_{\iota'}\rangle 
a^\ast(f_\iota)a(f_{\iota'}),
$$
holds on $\Gamma_-(\fh)$. If $q$ is trace class (in particular, if $q$ is finite 
rank), then $\d\Gamma(q)$ is bounded and belongs to $\CAR(\fh)$.

A unitary operator $u$ on $\fh$ induces a Bogoliubov automorphism of $\CAR(\fh)$
$$
A\mapsto \gamma_u(A)=\Gamma(u)A\Gamma(u)^\ast,
$$
such that $\gamma_u(a^\#(f))=a^\#(uf)$. If $t\mapsto u_t$ is a strongly 
continuous family of unitary operators on $\fh$, then $t\mapsto\gamma_{u_t}$ is a 
strongly continuous family of Bogoliubov automorphisms of $\CAR(\fh)$. In 
particular, if $u_t=\e^{\i tp}$ for some self-adjoint operator $p$ on $\fh$, we 
call $\gamma_{u_t}$ the quasi-free dynamics generated by $p$. 

The quasi-free dynamics generated by the identity $I$ is the gauge group of 
$\CAR(\fh)$ and  $N=\d\Gamma(I)$ is the number operator on $\Gamma_-(\fh)$,
$$
\vartheta^t(a^\#(f))=\e^{\i tN}a^\#(f)\e^{-\i tN}=a^\#(\e^{\i t}f)=\left\{
\begin{array}{rl}
\e^{-\i t}a(f)&\text{for }a^\#=a;\\[4pt]
\e^{\i t}a^\ast(f)&\text{for }a^\#=a^\ast.
\end{array}
\right.
$$
The algebra of observables of the Fermi gas is the gauge-invariant subalgebra of 
$\CAR(\fh)$,
$$
\CARv(\fh)=\{A\in\CAR(\fh)\,|\,\vartheta^t(A)=A\text{ for all }t\in\rr\}.
$$
It is the $C^\ast$-algebra generated by the set of all monomials in the $a^\#$ 
containing an equal number of $a$ and $a^\ast$ factors. In particular, it
is contained in the so-called even part of $\CAR(\fh)$ which consists of
all elements invariant under the involutive morphism $\Theta$ which maps 
$a^\#(f)$ to $-a^\#(f)$.

\subsubsection{Locally interacting dynamics}
The quasi-free dynamics generated by $h$  describes the sample coupled to the 
leads and $H=\d\Gamma(h)$ is the corresponding many-body Hamiltonian
$$
\tau_{H}^t(a^\#(f))=\e^{\i tH}a^\#(f)\e^{-\i tH}=a^\#(\e^{\i th}f).
\label{tauHdef}
$$
The group $\tau_{H}$ commutes with the gauge group $\vartheta$ so that it leaves 
$\CARv(\fh)$ invariant. In the following, we shall consistently denote 
one-particle operators with lower-case letters and capitalize the corresponding 
second quantized operator, e.g., $H_\cS=\d\Gamma(h_\cS)$, 
$H_\cR=\d\Gamma(h_\cR)$, etc. We shall also denote the corresponding groups of 
automorphism by $\tau_{H_\cS}$, $\tau_{H_\cR}$, etc.

For $x\in\cS$ we denote by $|x\rangle=\delta_x\in\fh_\cS$ the Kronecker delta at 
$x$ and by $a^\#_x=a^\#(\delta_x)$ the corresponding creation/annihilation
operators. We allow for interactions between particles in the sample $\cS$. 
However, particles in the leads remain free. The interaction energy within the 
sample is described by 
$$
W=\frac1{2}\sum_{x,y\in\cS}w(x,y)N_xN_y.
\label{Vint}
$$
where $N_x=a^\ast_xa_x$ and $w$ is a two-body potential satisfying 
$w(x,y)=w(y,x)\in\R$ and $w(x,x)=0$ for all $x,y\in\cS$.
For normalization purposes, we also assume that $\sup_{x,y\in\cS}|w(x,y)|=1$.
For any self-adjoint $W\in\CARv(\fh)$ and  any value of the interaction strength 
$\xi\in\rr$ the operator
$$
K=H+\xi W,
$$
is self-adjoint on the domain of $H$. 
Moreover $\tau_{K}^t(A)=\e^{\i tK}A\e^{-\i tK}$ defines a strongly continuous 
group of $\ast$-automorphisms of $\CAR(\fh)$ leaving invariant the subalgebra 
$\CARv(\fh)$. This group describes the full dynamics of the Fermi gas, including 
interactions. Another important operator is 
$$
K_D=H_{\cS}+\xi W+H_\cR,
$$
which describes the dynamics of the interacting but uncoupled system.
Note that $K=K_D+H_T$.

\subsubsection{States of the Fermi gas}\label{FermiState}
A state on $\CAR(\fh)$ is a linear functional
$$
\CAR(\fh)\ni A\mapsto\langle A\rangle\in\cc,
$$ 
such that $\langle A^\ast A\rangle\ge0$ for all $A$ and $\langle\one\rangle=1$. A 
state is gauge-invariant if $\langle\vartheta^t(A)\rangle=\langle A\rangle$ for 
all $t\in\rr$. Note that if $\st$ is a state on $\CAR(\fh)$ then its restriction 
to $\CARv(\fh)$ defines a state on this subalgebra. We shall use the same 
notation for this restriction. 

A state $\st$ on $\CAR(\fh)$ induces a GNS representation $(\cH,\pi,\Omega)$
where $\cH$ is a Hilbert space, $\pi$ is a $\ast$-morphism from $\CAR(\fh)$ to 
the bounded linear operators on $\cH$ and $\Omega\in\cH$ is a unit vector such 
that $\pi(\CAR(\fh))\Omega$ is dense in $\cH$ and 
$\langle A\rangle=(\Omega|\pi(A)\Omega)$ for all $A\in\CAR(\fh)$. Let $\rho$ be a 
density matrix on $\cH$ (a non-negative, trace class operator with $\tr(\rho)=1$).
The map $A\mapsto\tr(\rho\pi(A))$ defines a state on $\CAR(\fh)$. Such a state is 
said to be normal w.r.t.\;$\st$. From the thermodynamical point of view 
$\st$-normal states are close to $\st$ and describe local perturbations of this 
state.

Given a self-adjoint operator $\varrho$ on $\fh$ satisfying 
$0\le\varrho\le I$, the formula
$$
\langle a^\ast(f_1)\cdots a^\ast(f_k)a(g_l)\cdots a(g_1)\rangle_\varrho
=\delta_{kl}\det\{\langle g_j|\varrho| f_i\rangle\},
\label{GIQF}
$$
defines a unique gauge-invariant state on $\CAR(\fh)$. This state is called the 
quasi-free state of density $\varrho$. It is uniquely determined by the two point 
function $\langle a^\ast(f)a(g)\rangle_\varrho=\langle g|\varrho| f\rangle$. An 
alternative characterization of quasi-free states on $\CAR(\fh)$ is the usual 
fermionic Wick theorem
$$
\langle\varphi(f_1)\cdots\varphi(f_k)\rangle_\varrho=\left\{
\begin{array}{ll}
0,&\text{if $k$ is odd;}\\[8pt]
\ds\sum_{\pi\in\mathcal{P}_k}\varepsilon(\pi)
\prod_{j=1}^{k/2} \langle\varphi(f_{\pi(2j-1)})
\varphi(f_{\pi(2j)})\rangle_\varrho,
&\text{if $k$ is even};
\end{array}
\right.
\label{wickform}
$$
where $\varphi(f)=2^{-1/2}(a^\ast(f)+a(f))$ is the field operator, 
$\mathcal{P}_k$ denotes the set of pairings of $k$ objects, i.e., permutations 
satisfying $\pi(2j-1)<\min(\pi(2j),\pi(2j+1))$ for $j=1,\ldots,k/2$, and 
$\epsilon(\pi)$ is the signature of the permutation $\pi$.

Given a strongly continuous group $\tau$ of $\ast$-automorphisms of $\CAR(\fh)$ 
commuting with the gauge group $\vartheta$, a state $\st$ is a thermal 
equilibrium state at inverse temperature $\beta$ and chemical potential $\mu$ if 
it satisfies the $(\beta,\mu)$-KMS condition
w.r.t.\;$\tau$, i.e., if for any $A,B\in\CAR(\fh)$ the function
$$
F_{A,B}(s)=\langle A\tau^s\circ\vartheta^{-\mu s}(B)\rangle,
$$
has an analytic continuation to the strip $\{0<\Im s<\beta\}$ with a bounded 
continuous extension to the closure of this strip satisfying
\beq
F_{A,B}(s+\i\beta)=\langle\tau^s\circ\vartheta^{-\mu s}(B)A\rangle.
\label{KMScond}
\eeq
We shall say that such a state is a $(\beta,\mu)$-KMS state for $\tau$. 

\begin{remark} It is well known that for any $\beta>0$ and $\mu\in\rr$ the KMS
states $\st_{H}^{\beta,\mu}$ and $\st_{K}^{\beta,\mu}$ are thermodynamic limits
of the familiar grand canonical Gibbs states associated with the restrictions of
the Hamiltonian $H$ and $K$ to finitely extended reservoirs with appropriate
boundary conditions. See~\cite{BR2} for details. 
\end{remark}

\subsubsection{The initial state}
Let $\bbeta=(\beta_1,\ldots,\beta_m)\in\rr_+^m$, 
$\bmu=(\mu_1,\ldots,\mu_m)\in\rr^m$ and denote by 
$\langle\,\cdot\,\rangle_{H_j}^{\beta_j,\mu_j}$ the unique
$(\beta_j,\mu_j)$-KMS state for $\tau_{H_j}$ on $\CAR(\fh_j)$.
We say that a state $\st$ on $\CAR(\fh)$ is 
$(\bbeta,\bmu)$-KMS if it is a product state extension of the 
$\langle\,\cdot\,\rangle_{H_j}^{\beta_j,\mu_j}$ and some
gauge-invariant state $\langle\,\cdot\,\rangle_\cS$ on $\CAR(\fh_\cS)$,
i.e., if
$$
\langle A_1A_2\cdots A_mA_\cS\rangle
=\langle A_\cS\rangle_\cS
\prod_{j=1}^m\langle A_j\rangle_{H_j}^{\beta_j,\mu_j}
$$
holds for all $A_j\in\CAR(\fh_j)$ and $A_\cS\in\CAR(\fh_\cS)$.
We note that given a state $\st_\cS$ on $\CAR(\fh_\cS)$, such a
product state extension exists and is unique (see~\cite{AM}).
It describes a sample $\cS$ whose state $\st_\cS$ is unentangled
from the reservoirs which are all in thermal equilibrium.
In particular, a $(\bbeta,\bmu)$-KMS 
state needs not be quasi-free. We shall need the following extension
of the KMS condition~\eqref{KMScond} which follows from the CAR~\eqref{CAR},
the linearity, continuity and gauge invariance of states, the totality of 
monomials of the form
$A_1\cdots A_mA_\cS$ with $A_j\in\CAR(\fh_j)$ and $A_\cS\in\CAR(\fh_\cS)$ in 
$\CAR(\fh)$, and the KMS condition~\eqref{KMScond}:
For any $A\in\CAR(\fh)$ and $B\in\CAR(\fh_j)$ the function
$$
F_{A,B}(s)=\langle A\tau_{H_j}^s\circ\vartheta^{-\mu_js}(B)\rangle
$$
has an analytic extension to the strip $\{0<\Im s<\beta_j\}$ with a bounded
continuous extension to the closure of this strip satisfying
$F_{A,B}(s+\i\beta_j)=\langle\tau_{H_j}^s\circ\vartheta^{-\mu_js}(B)A\rangle$ 
or 
$$
\langle A\tau_{H_j}^s\circ\vartheta^{-\mu_js}(B)\rangle
=\langle\tau_{H_j}^{s-\i\beta_j}\circ\vartheta^{-\mu_j(s-\i\beta_j)}(B)A\rangle
$$ 
for all $s\in\rr$. In particular, when $s=0$, for any $A\in\CAR(\fh)$ and 
$B=a^\#(f)$ with $f\in \fh_j$ we have the identity:
\beq
\langle A \; a^\#(f)\rangle =\langle 
a^\#(\e^{\beta_j(h_j-\mu_j)}f)\; A\rangle.
\label{multiKMS}
\eeq
We also set
\begin{align}\label{iuni-3}
\varrho_\cR^{\bbeta,\bmu}
:=\bigoplus_j\left(I+\e^{\beta_j(h_j-\mu_j)}\right)^{-1},
\end{align}
which can also be seen as an operator on the whole $\fh$ by extending it by zero 
on $\fh_S$.  

\section{Main results}\label{mainresults}
In this section we introduce the main objects of interest for the
NEGFs machinery and we state our main results.

\subsection{Retarded and advanced Green's Functions}
For motivation purposes, let us first consider the non-interacting case, i.e., 
set $\xi=0$. The one-body wave function then satisfies the Schr\"odinger equation
$$
\i\partial_t\varphi(t)=h\varphi(t),
$$
and the associated Cauchy problem with initial condition 
$\varphi(t_0)=\varphi_0\in\fh$
is solved by the unitary propagator $\varphi(t)=\e^{-\i(t-t_0)h}\varphi_0$.
In order to study the response of the system to time-dependent perturbations
one investigates the corresponding inhomogeneous equation
$$
(\i\partial_s-h)\varphi(s)=\psi(s).
\label{InhomSch}
$$
To deal with perturbations that are localized in time, it makes sense to consider 
this equation in the Hilbert space $\cH=L^2(\rr,\d s;\fh)$. The Fourier transform
\[
\hat{\varphi}(\omega)=\int_\rr\varphi (s)\e^{-\i\omega s}\d s,
\]
maps the time-domain Hilbert space $\cH$ unitarily onto the frequency-domain 
Hilbert space $L^2(\rr,\frac{\d\omega}{2\pi};\fh)$ in such a way that
$$
((\i\partial_s-h)\varphi)\widehat{\ }(\omega)=(-\omega-h)\hat{\varphi}(\omega).
$$
Thus, the operator $\Omega=\i\partial_s-h$, with domain
\begin{align}\label{iuni-1}
\cH^1=H^1(\rr,\d s;\fh)=\{\varphi\in\cH\,|\,
\|\varphi\|_1=\|\sqrt{1+\omega^2}\hat{\varphi}\|<\infty\}
\end{align}
is self-adjoint on $\cH$. Its spectrum fills the real axis and a simple 
calculation shows that the unitary group it generates is given by
$$
(\e^{\i t\Omega}\varphi)(s)=\e^{-\i th}\varphi(s-t).
$$
It follows that for $z\in\cc_\pm=\{z\in\cc\,|\,\pm\Im z>0\}$ the resolvent
$$
G_0^\pm(z)=(\Omega-z)^{-1}=\pm\i\int_{\rr_\pm}\e^{-\i t(\Omega-z)}\d t,
$$
has the time-domain expression
\beq\label{aprili2}
(G_0^\pm(z)\varphi)(s)
=\pm\i\int_\rr\theta(\pm(s'-s))\e^{\i(s'-s)(h+z)}\varphi(s')\d s',
\eeq
where $\theta$ denotes Heaviside step function. 

We set $\rr_\pm=\{s\in\rr\,|\,\pm s>0\}$ and observe that
the boundary values $G_0^\pm(E)=G_0(E\pm\i0)$ are well defined as
operators on the Fr\'echet space $\cH_{\loc\mp}=L^2_\loc(\rr_\mp,\d s;\fh)$,
with the semi-norm estimate
$$
\|G_0^\pm(E)\varphi\|_T\le T\|\varphi\|_T,
$$
where $\|\varphi\|_T^2=\int_0^T\|\varphi(\mp s)\|^2\d s$.  The 
`integral kernel' of the operator $G_0^\mp(E):\cH_{\loc\pm}\to\cH_{\loc\pm}$ 
is called retarded/advanced Green's function. In the physics literature it is 
usually denoted by $G_0^R(E|s,s')/G_0^A(E|s,s')$ or simply 
$G_0^R(s,s')/G_0^A(s,s')$ in the special case $E=0$. Observing 
that for $f,g\in\fh$
\begin{align*}
\langle f|G_0^R(E|s,s')|g\rangle
=-\i\theta(s-s')\e^{\i(s'-s)E}\left\langle\left\{\tau_H^{s'}(a^\ast(g)),
\tau_H^s(a(f))\right\}\right\rangle,\\[4pt]
\langle f|G_0^A(E|s,s')|g\rangle
=+\i\theta(s'-s)\e^{\i(s'-s)E}\left\langle\left\{\tau_H^{s'}(a^\ast(g)),
\tau_H^s(a(f))\right\}\right\rangle,
\label{freeGAR}
\end{align*}
leads to the definition of the retarded/advanced interacting Green's function
\beq
\begin{split}
\langle f|G^R(E|s,s')|g\rangle
:=-\i\theta(s-s')\e^{\i(s'-s)E}\left\langle\left\{\tau_K^{s'}(a^\ast(g)),
\tau_K^s(a(f))\right\}\right\rangle,\\[4pt]
\langle f|G^A(E|s,s')|g\rangle
:=+\i\theta(s'-s)\e^{\i(s'-s)E}\left\langle\left\{\tau_K^{s'}(a^\ast(g)),
\tau_K^s(a(f))\right\}\right\rangle.
\end{split}
\label{intGAR}
\eeq
The decoupled retarded/advanced Green's function $G_D^{A/R}$ are defined similarly
by replacing $\tau_K$ by $\tau_{K_D}$ in~\eqref{intGAR}. We observe that
for $f,g\in\fh_\cR$, the CAR and the fact that there is no interaction 
in the reservoirs imply that
\begin{align*}
\langle f|G_D^R(s,s')|g\rangle
&=-\i\theta(s-s')\langle f| \e^{\i(s'-s)h_\cR}|g\rangle,\\[4pt]
\langle f|G_D^A(s,s')|g\rangle
&=+\i\theta(s'-s)\langle f|\e^{\i(s'-s)h_\cR}|g\rangle.
\label{2014-mai-4}
\end{align*}

\subsection{Other Green's Functions}

For $s,s'\geq 0$ and $f,g\in\fh$, the interacting `lesser' and `greater' Green's
functions are defined by
\beq
\begin{split}
\langle f|G^<(s,s')|g\rangle&:=+\i\left\langle\tau_K^{s'}(a^*(g))
\tau_K^{s}(a(f))\right\rangle,\\[4pt]
\langle f|G^>(s,s')|g\rangle&:=-\i\left\langle\tau_K^{s}(a(f))
\tau_K^{s'}(a^*(g))\right\rangle,
\end{split}
\label{2014-mai-2}
\eeq
and play an important role in the NEGF formalism. The decoupled `lesser' and 
`greater' Green's functions are obtained upon replacement of $K$ by $K_D$ in
the above expressions. For $f,g\in\cH_\cR$,  the fact that the restriction of
the state $\st$ to $\CAR(\fh_\cR)$ is the gauge-invariant quasi-free state with 
density $\varrho_\cR^{\bbeta,\bmu}$ leads to the formulas
\begin{align*}
\langle f|G_D^<(s,s')|g\rangle
&=+\i\langle f|\varrho_\cR^{\bbeta,\bmu}\e^{\i(s'-s)h_\cR}|g\rangle,\\[4pt]
\langle f|G_D^>(s,s')|g\rangle
&=-\i\langle f|(I-\varrho_\cR^{\bbeta,\bmu})\e^{\i(s'-s)h_\cR}|g\rangle.
\end{align*}

For completeness, let us mention two combinations of the
`lesser' and  `greater' Green's functions which also appear in the physics
literature. We should not, however, use them in the following.
The `spectral function' is
$$
\langle f|A(s,s')|g\rangle:=\i\langle f|G^R(s,s')-G^A(s,s')|g\rangle
=\i\langle f|G^>(s,s')-G^<(s,s')|g\rangle=\left\langle\left\{
\tau_K^{s'}(a^*(g)),\tau_K^{s}(a(f))\right\}\right\rangle,
$$
with the property that $A(t,t)=I$, while the `Keldysh' Green's function 
is
$$
\langle f|G^K(s,s')|g\rangle
:=\langle f|G^<(s,s')+G^>(s,s')|g\rangle=\left\langle\i\left[
\tau_K^{s'}(a^*(g)),\tau_K^{s}(a(f))\right]\right\rangle.
$$

\subsection{The Jauho-Meir-Wingreen current formula} 
From Eq.~\eqref{2014-mai-2} we see that the interacting lesser 
Green's function $G^<$ encodes all one-particle properties of the system. 
For example, the sample's particle density at time $t\ge0$ is given by
$$
\varrho(x,t)
=\langle\tau_K^t(a^\ast_xa_x)\rangle
=\Im\langle x|G^<(t,t)|x\rangle.
$$
Computing the time derivative of the sample's particle number 
$N_\cS=\d\Gamma(1_\cS)$ we obtain
\begin{align*}
\partial_t\tau_K^t(N_\cS)=\tau_K^t(\i[K,N_\cS])
&=\tau_K^t(\i[H_T,N_\cS])\\
&=\sum_jd_j\tau_K^t(\i[\d\Gamma(|\phi_j\rangle\langle\psi_j|
+|\psi_j\rangle\langle\phi_j|),\d\Gamma(1_\cS)])\\
&=\sum_jd_j\tau_K^t(\d\Gamma(\i[|\phi_j\rangle\langle\psi_j|
+|\psi_j\rangle\langle\phi_j|,1_\cS]))
=\sum_jJ_j(t)
\end{align*}
which allows us to identify the \jth term in the above sum
$$
J_j(t)=\i d_j\tau_K^t\left(
a^*(\psi_j)a(\phi_j)-a^*(\phi_j)a(\psi_j)
\right),
$$
with the particle current out of the \jth reservoir.
Its expectation value in the initial state is 
\beq\label{curentt1}
I_j(t):=\langle J_j(t)\rangle
=d_j\left(\langle\phi_j|G^<(t,t)|\psi_j\rangle
-\langle\psi_j|G^<(t,t)|\phi_j\rangle\right)
=2d_j\Re\langle\phi_j|G^<(t,t)|\psi_j\rangle,
\eeq
where, in the last equality, we used the fact that 
$G^<(t,t)^\ast=-G^<(t,t)$.
The main result of this paper is a rigorous proof of the JMW formula:

\begin{theorem}[The Jauho-Meir-Wingreen formula]\label{jauho}
If the initial state $\st$ is $(\bbeta,\bmu)$-KMS, then the 
particle current out of the \jth reservoir at time $t>0$ is given by
$$
I_j(t)=-2d_j^2\int_0^t\d s\int\d\nu_j(E)\,
{\rm Im}\left \{\e^{\i(t-s)E}\langle\phi_j|G^<(t,s)
+G^R(t,s)[1+\e^{\beta_j(E-\mu_j)}]^{-1}|\phi_j\rangle\right\},
$$
where $\nu_j$ denotes the spectral measure of $h_j$ for the vector $\psi_j$.
\end{theorem}

\begin{remark}
The main feature of the JMW formula is that it only involves
interacting Green functions restricted to the sample $\cS$ (in spite of the fact that in~\eqref{curentt1} $\phi_j\in \fh_\cS$ and 
$\psi_j\in\fh_j$). If one is interested in 
transient regimes, it seems that this formula is easier to deal with from a 
numerical point of view. But if one is interested in proving the convergence 
to a steady state value when $t\to\infty$, it is not very useful. Moreover, the JMW formula must be backed-up by systematic methods of calculating
interacting GFs, which rely on Dyson equations and interaction self-energies.
\end{remark}
\begin{remark}
Let $\widetilde{\fh}_j\subset\fh_j$ denote the cyclic subspace of $h_j$
generated by $\psi_j$, i.e., the smallest $h_j$-invariant closed subspace
of $\fh_j$ containing $\psi_j$. Set 
$\tilde{\fh}=\fh_\cS\oplus(\oplus_j)\tilde{\fh}_j$ and denote by 
$\tilde{\fh}^\perp$ the orthogonal complement of $\tilde{\fh}$ in $\fh$. 
Let $\cO$ be the $\ast$-subalgebra of $\CAR(\fh)$ generated by 
$\{a(f)\,|\,f\in\tilde{\fh}\}$. One easily checks that $\cO$ is invariant
under the groups $\vartheta$ and $\tau_K$. By the exponential law for 
fermions (see, e.g., Section 3.4.3 in~\cite{DeGe}) there exists a unitary map
$U:\Gamma_-(\fh)\to\Gamma_-(\tilde{\fh})\otimes\Gamma_-(\tilde{\fh}^\perp)$
such that $Ua(f)U^\ast=a(f)\otimes I$ for $f\in\tilde{\fh}$. 
Identifying in this way $\cO$ with $\CAR(\tilde{\fh})$ and noticing that 
$W\in\cO$, one easily shows that the restriction of the group $\tau_K$ to 
$\cO$ is the Heisenberg dynamics generated by 
$\tilde K=\d\Gamma(\tilde h)+\xi W$ where $\tilde h$ is the
restriction of $h$ to $\tilde{\fh}$. Thus, as far as the dynamics
on $\cO$ is concerned, we may assume w.l.o.g.\;that $\psi_j$ is a cyclic
vector for $h_j$. Going to the induced spectral representation, this means
that $\fh_j=L^2(\rr,\d\nu_j(E))$ where $\nu_j$ (the spectral measure of
$h_j$ for $\psi_j$) is a probability measure, $h_j$ is multiplication by $E$ 
and $\psi_j$ is the constant function $\psi_j(E)=1$.
\end{remark}

Theorem~\ref{jauho} is a direct consequence of Eq.~\eqref{curentt1} and
the next result.

\begin{proposition}[The Langreth identity]\label{langreth}
Assume that the initial state $\st$ is $(\bbeta,\bmu)$-KMS.
Then, the following identity holds for all $j\in\{1,\ldots,m\}$ and 
$t,t'\geq 0$,
\begin{align}\label{2014-mai-7}
\langle\phi_j|G^<(t,t')|\psi_j\rangle=d_j\int_0^\infty
\langle\phi_j|\left(G^R(t,s)|\phi_j\rangle\langle\psi_j|G_D^<(s,t')
+G^<(t,s)|\phi_j\rangle\langle\psi_j|G_D^A(s,t')\right)|\psi_j\rangle
\d s.
\end{align} 
\end{proposition}

\begin{remark}
 One should compare our results with formulas~(12.11), (12.19), (12.21) and~(13.3)
in~\cite{HJ}. Our formula~\eqref{2014-mai-7} corresponds to  (12.19) and (13.3) in~\cite{HJ}. Haug and Jauho derive these two formulas in two steps: first using Keldysh contours (see (12.18)
in~\cite{HJ}) and after that a Langreth-type `analytic continuation' in order to
come back to `normal' integrals.
\end{remark}

\subsection { A decoupling Keldysh-like identity}
As one can see from the JMW formula, one is left with computing correlation
functions between points both situated in the sample $\cS$. From a mathematical
point of view, this is as complicated as the original problem. Nevertheless, in
the physics literature one tries to rewrite the interacting lesser functions
in terms of retarded/advanced GFs which afterward can be numerically computed 
by solving Dyson type equations. The next proposition shows how this is done:

\begin{proposition}[A decoupling Keldysh identity]\label{Dkeldysh}
If the initial state $\st$ is $(\bbeta,\bmu)$-KMS and if its
restriction to $\CAR(\fh_\cS)$ is the vacuum state, then there exists a 
continuous function 
$$
\rr_+\times\rr_+\ni
(s,s')\mapsto S^<(s,s')=\sum_{x,x'\in\cS}|x\rangle S_{xx'}^<(s,s')\langle x'|
\in\cB(\fh_\cS)
$$
such that
\beq
\label{iulie140}
\langle\phi|G^<(t,t')|\phi'\rangle
=\int_0^\infty\d s\int_0^\infty\d s'\,
\langle\phi|G_0^R(t,s)S^<(s,s')G_0^A(s',t')|\phi'\rangle,
\eeq
for $\phi,\phi'\in\fh_\cS$ and $t,t'\ge0$. Moreover,
$$
S_{xx'}^<(s,s')=\i\langle\cT^\ast_{x'}(s')\cT_x(s)\rangle
$$
where
$$
\cT_x(s)=a(\e^{\i sh_D}h_T\delta_x)+\xi\tau_K^s(a_xV_x).
$$
and
$$
V_x=\sum_{y\in\cS}w(x,y)N_y.
$$
As a function of the interaction strength $\xi$, $S^<(s,s')$ is
entire analytic. The first two terms of its expansion
$$
S^<(s,s')=\sum_{n\ge0}\xi^nS^{<(n)}(s,s')
$$
are given by (recall~\eqref{iuni-3} for the definition of 
$\varrho_\cR^{\bbeta,\bmu}$):
$$
S^{<(0)}_{xx'}(s,s')=\i\langle x|h_T\e^{-\i sh_\cR}
\varrho_\cR^{\bbeta,\bmu}
\e^{\i s'h_\cR}
h_T|x'\rangle=\i\sum_jd_j^2
\left(\int\frac{\e^{\i(s'-s)E}}{1+\e^{\beta_j(E-\mu_j)}}\d\nu_j(E)\right)
\langle x|\phi_j\rangle\langle\phi_j|x'\rangle,\\
$$
and
\begin{align*}
S^{<(1)}_{xx'}(s,s')=\i\sum_{y\in\cS}\biggl[
\phantom{+}w(x,y)\biggl(
&\langle x|\e^{-\i sh}\varrho_\cR^{\bbeta,\bmu}\e^{\i sh}|y\rangle
\langle y|\e^{-\i sh}\varrho_\cR^{\bbeta,\bmu}\e^{\i s'h_\cR}h_T|x'\rangle\\
-&\langle x|\e^{-\i sh}\varrho_\cR^{\bbeta,\bmu}\e^{\i s'h_\cR}h_T|x'\rangle
\langle y|\e^{-\i sh}\varrho_\cR^{\bbeta,\bmu}\e^{\i sh}|y\rangle\biggr)\\
+w(x',y)\biggl(&\langle x|h_T
\e^{-\i sh_\cR}\varrho_\cR^{\bbeta,\bmu}\e^{\i s'h}|y\rangle
\langle y|\e^{-\i s'h}\varrho_\cR^{\bbeta,\bmu}
\e^{\i s'h}|x'\rangle\\
-&\langle x|h_T
\e^{-\i sh_\cR}\varrho_\cR^{\bbeta,\bmu}\e^{\i s'h}|x'\rangle
\langle y|\e^{-\i s'h}\varrho_\cR^{\bbeta,\bmu}
\e^{\i s'h}|y\rangle\biggr)\biggr].
\end{align*}
\end{proposition}
The `true' Keldysh identity appearing in the physics literature differs
from~\eqref{iulie140} in that it involves the interacting retarded/advanced
GF's, $G^{A/R}$, instead of the non-interacting ones $G_0^{A/R}$. The argument
leading to such a relation relies on a Dyson type equation connecting $G^{A/R}$
to $G_0^{A/R}$ through the so-called advanced/retarded self-energies. In the
physics literature, the derivation of these Dyson equations usually rests on
formal analogies with zero temperature and/or diagrammatic perturbative
techniques for the contour ordered GFs. The next two results stated in
Proposition~\ref{Dyson} and Proposition~\ref{Ikeldysh} are non-perturbative and
provide a rigorous mathematical foundation to these relations.

\subsection{Dyson equations}
Given the construction in the beginning of this section relating the retarded
and advanced Green's functions to boundary values of resolvent operators, it is
clear that the Green's functions $G_v^{R/A}$ of a time-independent
perturbation $v$ of the one-body Hamiltonian, where $v\in\cB(\cH)$, are related
to the unperturbed Green's functions $G_0^{R/A}$ by the second resolvent
equation, i.e.,
\begin{align*}
G^R_{ v}(E|s,s')&=G_0^R(E|s,s')
+\int_{s'}^{s}G_0^R(E|s,r)\;{ v}\;G^R_{ v}(E|r,s')\d r,\\[4pt]
G_v^A(E|s,s')&=G_0^A(E|s,s')
+\int_s^{s'}G_0^A(E|s,r)\;v\; G_v^A(E|r,s')\d r,
\end{align*}
which we write as
\beq
G_v^{R/A}=G_0^{R/A}+G_0^{R/A}vG_v^{R/A}=G_0^{R/A}+G_v^{R/A}vG_0^{R/A}.
\label{Dysonv}
\eeq
We now state our second result concerning the existence of interacting proper
self-energies and formulate the Dyson equations, similar to~\eqref{Dysonv},
obeyed by the interacting advanced/retarded Green functions.

\begin{proposition}[The advanced/retarded Dyson equation]\label{Dyson}
Let $\st$ be an arbitrary state on $\CAR(\fh)$. There exists continuous
functions 
$$
\rr_+\times\rr_+\ni
(s,s')\mapsto
\Sigma^{A/R}(s,s')=\sum_{x,x'\in\cS}
|x\rangle
\Sigma^{A/R}_{xx'}(s,s')\;
\langle x'|\in\cB(\fh_\cS)
$$
such that for every $s,s'\geq 0$ we have: 
\begin{align}\label{aprili1}
G^{A/R}(E|s,s')&=G_0^{A/R}(E|s,s')+
{ \int_0^\infty}\d r{ \int_0^\infty}\d r'\,
G_0^{A/R}(E|s,r)\Sigma^{A/R}(E|r,r')G^{A/R}(E|r',s')\\[4pt]
&=G_0^{A/R}(E|s,s')+
{ \int_0^\infty}\d r{ \int_0^\infty}\d r'\,
G^{A/R}(E|s,r)\Sigma^{A/R}(E|r,r')G_0^{A/R}(E|r',s').\nonumber
\end{align}
Moreover, the map $\xi\mapsto\Sigma^{A/R}\in\cB(\fh_\cS)$  which defines the irreducible advanced/retarded self-energy is entire analytic.
\end{proposition}

\subsection{The Keldysh identity}
We note an important feature of the above Dyson equations. Due to the special
structure of the self-energy $\Sigma^{A/R}$ (which only lives in the sample), we
see that the finite dimensional restriction of $G^{A/R}(E|s,s')$ to the small
sample obeys the same equation.

Also, by isolating $G_0^{A/R}$ from~\eqref{aprili1} and introducing it
in~\eqref{iulie140} we immediately obtain the following result:

\begin{proposition}[The Keldysh identity]\label{Ikeldysh}
If the initial state $\st$ is $(\bbeta,\bmu)$-KMS and if its
restriction to $\CAR(\fh_\cS)$ is the vacuum state,
then there exists a continuous function 
$$
\R_+\times \R_+\ni (s,s')\mapsto {\Sigma}^<(s,s')
=\sum_{x,x'\in \cS}|x\rangle{\Sigma}_{xx'}^<(s,s')\; \langle x'|\in\cB(\fh_\cS)
$$ such that
\beq
\label{iulie1401}
\langle\phi|G^<(t,t')|\phi'\rangle
=\int_0^\infty\d s\int_0^\infty\d s'
\langle\phi|G^R(t,s){\Sigma}^<(s,s')G^A(s',t')|\phi'\rangle,
\eeq
for $\phi,\phi'\in\fh_\cS$ and $t,t'\ge0$.
\end{proposition}

\section{Proofs}\label{prooofs}
\subsection{Proof of Proposition \ref{langreth}}
One main ingredient of the proof is the following Duhamel formula
\beq
\tau_K^{t'}(A)
=\tau_{K_D}^{t'}(A)+\int_0^{t'}\tau_K^{s}(\i[H_T,\tau_{K_D}^{t'-s}(A)])\,\d s,
\label{Duhamel0}
\eeq
which is obtained by differentiating $\tau_K^{-s}\circ\tau_{K_D}^{s}(A)$, then 
integrating back from $0$ to $t'$, and finally changing the integration variable 
from $s$ to $t'-s$. Applying it to $A=a^\ast(\psi_j)$ we infer
\beq
\langle\phi_j|G^<(t,t')|\psi_j\rangle
=\i\left\langle\tau_{K_D}^{t'}(a^\ast(\psi_j))
\tau_K^t(a(\phi_j))\right\rangle
+\int_0^{t'}\i\left\langle\tau_{K}^{s}(
\i[H_T,\tau_{K_D}^{t'-s}(a^\ast(\psi_j))])
\tau_K^t(a(\phi_j))\right\rangle\d s.
\label{Duhamel1}
\eeq
For $t'-s\ge0$ we have
\begin{align*}
\i[H_T,\tau_{K_D}^{t'-s}(a^\ast(\psi_j))]
&=\i[\d\Gamma(h_T),a^\ast(\e^{\i(t'-s)h_j}\psi_j)]
=a^\ast(\i h_T\e^{\i(t'-s)h_j}\psi_j)\\
&=\i \,d_j\langle\psi_j|\e^{\i(t'-s)h_j}|\psi_j\rangle a^\ast(\phi_j)
=d_j\langle\psi_j|G_D^A(s,t')|\psi_j\rangle a^\ast(\phi_j),
\end{align*}
and~\eqref{Duhamel1} becomes
\beq
\langle\phi_j|G^<(t,t')|\psi_j\rangle
=\i\left\langle\tau_{K_D}^{t'}(a^\ast(\psi_j))
\tau_K^t(a(\phi_j))\right\rangle
+d_j\int_0^\infty 
\langle\phi_j|G^<(t,s)|\phi_j\rangle\langle\psi_j|G_D^A(s,t')|\psi_j\rangle\d s.
\label{halfofit}
\eeq
We see that the second term on the right hand side coincides with the second term 
on the rhs of~\eqref{2014-mai-7}. The rest of this proof will deal with the first 
term on the rhs of~\eqref{halfofit}.

Since $h_T$ is finite rank, $H_T=\d\Gamma(h_T)$ is a self-adjoint element of 
$\CAR(\fh)$. This allows us to introduce the {\sl interaction representation} of 
the full dynamics: for any $A\in\CAR(\fh)$ and $t\in\rr$,
$$
\tau_K^t(A)=\Gamma_t^\ast\tau_{K_D}^t(A)\Gamma_t,
$$
where the cocycle $\Gamma_t=\e^{\i tK_D}\e^{-\i tK}$ satisfies the Cauchy problem
$$
\i\partial_t\Gamma_t=\tau_{K_D}^t(H_T)\Gamma_t,\qquad\Gamma_0=\one,
$$
and takes its values in the unitary elements of $\CAR(\fh)$. It follows that
\beq
\i\left\langle\tau_{K_D}^{t'}(a^\ast(\psi_j))
\tau_K^t(a(\phi_j))\right\rangle
=\i\left\langle\tau_{K_D}^{t'}(a^\ast(\psi_j))
\Gamma_t^\ast\tau_{K_D}^t(a(\phi_j))\Gamma_t\right\rangle
=\i\left\langle a^\ast(\e^{\i t'h_j}\psi_j)
\Gamma_t^\ast\tau_{K_D}^t(a(\phi_j))\Gamma_t\right\rangle.
\label{foo1}
\eeq
We shall now consider the expression $\i\langle a^\ast(\tilde{\psi}_j)
\Gamma_t^\ast\tau_{K_D}^t(a(\phi_j))\Gamma_t\rangle$
for arbitrary $\tilde{\psi}_j\in\fh_j$. Using the canonical
anti-commutation relations which implies
\beq
\{a^\ast(\tilde{\psi}_j),\tau_{K_D}^t(a(\phi_j))\}
=\tau_{K_D}^t(\{a^\ast(\e^{-\i th_j}\tilde{\psi}_j),a(\phi_j)\})=0,
\label{CARform}
\eeq
we commute the creation operator to the right, getting 
\begin{align}
\i\left\langle a^\ast(\tilde{\psi}_j)
\Gamma_t^\ast\right.&\left.%
\vphantom{\tilde{\psi}_j}\!%
\tau_{K_D}^t(a(\phi_j))\Gamma_t\right\rangle
=\i\left\langle 
\Gamma_t^\ast a^\ast(\tilde{\psi}_j)\tau_{K_D}^t(a(\phi_j))\Gamma_t
+[a^\ast(\tilde{\psi}_j),\Gamma_t^\ast]\tau_{K_D}^t(a(\phi_j))\Gamma_t
\right\rangle\nonumber\\
&=\i\left\langle 
-\Gamma_t^\ast\tau_{K_D}^t(a(\phi_j))a^\ast(\tilde{\psi}_j)\Gamma_t
+[a^\ast(\tilde{\psi}_j),\Gamma_t^\ast]\tau_{K_D}^t(a(\phi_j))\Gamma_t
\right\rangle\label{Commut1}\\
&=\i\left\langle 
-\Gamma_t^\ast\tau_{K_D}^t(a(\phi_j))\Gamma_t a^\ast(\tilde{\psi}_j)
-\Gamma_t^\ast\tau_{K_D}^t(a(\phi_j))[a^\ast(\tilde{\psi}_j),\Gamma_t]
+[a^\ast(\tilde{\psi}_j),\Gamma_t^\ast]\tau_{K_D}^t(a(\phi_j))\Gamma_t
\right\rangle.\nonumber
\end{align}
Since the initial state is $(\bbeta,\bmu)$-KMS, the KMS condition~\eqref{multiKMS}
implies
\begin{align*}
\left\langle\Gamma_t^\ast\tau_{K_D}^t(a(\phi_j))\Gamma_t 
a^\ast(\tilde{\psi}_j)\right\rangle
&=\left\langle a^\ast(\e^{\beta_j(h_j-\mu_j)}\tilde{\psi}_j)\Gamma_t^\ast 
\tau_{K_D}^t(a(\phi_j))\Gamma_t\right\rangle,
\end{align*}
which, combined with~\eqref{Commut1}, yields
$$
\i\left\langle a^\ast([1+\e^{\beta_j(h_j-\mu_j)}]\tilde{\psi}_j)
\Gamma_t^\ast\tau_{K_D}^t(a(\phi_j))\Gamma_t\right\rangle
=\i\left\langle 
-\Gamma_t^\ast\tau_{K_D}^t(a(\phi_j))[a^\ast(\tilde{\psi}_j),\Gamma_t]
+[a^\ast(\tilde{\psi}_j),\Gamma_t^\ast]\tau_{K_D}^t(a(\phi_j))\Gamma_t
\right\rangle.
$$
Setting 
$\tilde{\psi}_j=[1+\e^{\beta_j(h_j-\mu_j)}]^{-1}\e^{\i t'h_j}\psi_j$
we can thus rewrite~\eqref{foo1} as
$$
\i\left\langle\tau_{K_D}^{t'}(a^\ast(\psi_j))
\tau_K^t(a(\phi_j))\right\rangle
=\i\left\langle 
-\Gamma_t^\ast\tau_{K_D}^t(a(\phi_j))[a^\ast(\tilde{\psi}_j),\Gamma_t]
+[a^\ast(\tilde{\psi}_j),\Gamma_t^\ast]\tau_{K_D}^t(a(\phi_j))\Gamma_t
\right\rangle.
$$
We now undo the commutators and obtain four terms, two of them forming an anti-commutator equal to zero due to~\eqref{CARform}. This leads to
\beq
\i\left\langle\tau_{K_D}^{t'}(a^\ast(\psi_j))
\tau_K^t(a(\phi_j))\right\rangle
=\i\left\langle
\{a^\ast(\tilde{\psi}_j),\tau_K^t(a(\phi_j))\}
\right\rangle
=\i\left\langle\tau_K^t\left(
\{\tau_K^{-t}(a^\ast(\tilde{\psi}_j)),a(\phi_j)\}\right)
\right\rangle.
\label{Hungry}
\eeq
Another application of a slightly modified version of the Duhamel formula 
in~\eqref{Duhamel0} further gives: 
\begin{align*}
\{\tau_K^{-t}(a^\ast(\tilde{\psi}_j)),a(\phi_j)\}
&=\{\tau_{K_D}^{-t}(a^\ast(\tilde{\psi}_j)),a(\phi_j)\}
-\int_0^t\left\{\tau_K^{-(t-s)}(\i[H_T,\tau_{K_D}^{-s}(a^\ast(\tilde{\psi}_j))]),a(\phi_j)\right\}\d s\\
&=-d_j\int_0^t\{\tau_K^{-(t-s)}(a^\ast(\phi_j)),a(\phi_j)\}
\i\langle\psi_j|\e^{-\i sh_j}|\tilde{\psi}_j\rangle\d s\\
&=-d_j\int_0^t\{\tau_K^{-(t-s)}(a^\ast(\phi_j)),a(\phi_j)\}
\langle\psi_j|G_D^<(s,t')|\psi_j\rangle\d s,
\end{align*}
The anti-commutator on the rhs of the first equality equals zero as in~\eqref{CARform}.
Inserting this relation into~\eqref{Hungry} we finally get
\begin{align*}
\i\left\langle\tau_{K_D}^{t'}(a^\ast(\psi_j))
\tau_K^t(a(\phi_j))\right\rangle
&=-d_j\int_0^t\i\left\langle
\{\tau_K^{s}(a^\ast(\phi_j)),\tau_K^t(a(\phi_j))\}
\right\rangle\langle\psi_j|G_D^<(s,t')|\psi_j\rangle\d s\\
&=d_j\int_0^\infty\langle\phi_j|G^R(t,s)|\phi_j\rangle
\langle\psi_j|G_D^<(s,t')|\psi_j\rangle\d s,
\end{align*}
which, together with~\eqref{halfofit} yields the result.

\subsection{Proof of Proposition \ref{Dkeldysh}}
Recall that $N_\cS$ denotes the particle number operator of the sample $\cS$. 
For any $A\in\CAR(\fh)$ and any unit vector $\phi\in\fh_\cS$ one has
$$
|\left\langle a^\ast(\phi)A\right\rangle|
\le\left\langle a^\ast(\phi)a(\phi)\right\rangle^{1/2}
\left\langle A^\ast A\right\rangle^{1/2}
\le\left\langle N_\cS\right\rangle^{1/2}
\left\langle A^\ast A\right\rangle^{1/2}.
$$
Since in this proposition we assume that the restriction of 
$\left\langle\,\cdot\,\right\rangle$ to $\CAR(\fh_\cS)$ 
is the vacuum state, one has $\left\langle N_\cS\right\rangle=0$ and hence
$\left\langle a^\ast(\phi)A\right\rangle=0$ for all $A\in\CAR(\fh)$. In the 
following, we shall write $A\sim B$ whenever $A,B\in\CAR(\fh)$ are such that
$$
\left\langle(A-B)^\ast(A-B)\right\rangle=0,
$$
and hence $\left\langle CA\right\rangle=\left\langle CB\right\rangle$
for all $C\in\CAR(\fh)$. 

Starting with Duhamel's formula for the pair $(K,H)$:
$$
\tau_K^t(a(\phi))=\tau_H^t(a(\phi))
+\xi\int_0^t\tau_K^s\left(\i[W,\tau_H^{t-s}(a(\phi))]\right)\d s,
$$
we note that
\begin{align*}
\tau_H^t(a(\phi))&=\tau_{H_D}^t(a(\phi))
+\int_0^t\tau_{H_D}^s\left(\i[H_T,\tau_{H}^{t-s}(a(\phi))]\right)\d s\\
&=a(\e^{\i th_D}\phi)+
\int_0^ta(\e^{\i sh_D}\i h_T\e^{\i(t-s)h}\phi)\d s\\
&=a(\e^{\i th_D}\phi)+
\int_0^\infty\sum_jd_j\left(
\langle\phi|{ G}_0^R(t,s)|\psi_j\rangle a(\e^{\i sh_D}\phi_j)
+\langle\phi|{ G}_0^R(t,s)|\phi_j\rangle a(\e^{\i sh_D}\psi_j)\right)\d s,
\end{align*}
from which we deduce
$$
\tau_H^t(a(\phi))\sim\int_0^\infty\sum_jd_j
\langle\phi|{ G}_0^R(t,s)|\phi_j\rangle a(\e^{\i sh_D}\psi_j)\d s.
$$
Setting
$$
V_x:=\sum_{y\in\cS}w(x,y)N_y,
$$
we further write, for $t>s$,
\begin{align*}
\i[W,\tau_H^{t-s}(a(\phi))]
&=\frac12\sum_{x,y\in\cS}w(x,y)
\i[N_xN_y,a(\e^{\i(t-s)h}\phi)]\\
&=\frac12\sum_{x,y\in\cS}w(x,y)\left(
N_x\i[N_y,a(\e^{\i(t-s)h}\phi)]+{ \i}[N_x,a(\e^{\i(t-s)h}\phi)]N_y\right)\\
&=\frac12\sum_{x,y\in\cS}w(x,y)\left(
N_x\langle\phi|{ G}_0^R(t,s)|y\rangle a_y
+\langle\phi|{ G}_0^R(t,s)|x\rangle a_xN_y\right)\\
&=\sum_{x\in\cS}\langle\phi|{ G}_0^R(t,s)|x\rangle a_xV_x.
\end{align*}
Thus,
\begin{align*}
\tau_K^t(a(\phi))&\sim
\int_0^\infty\sum_jd_j
\langle\phi|{ G}_0^R(t,s)|\phi_j\rangle a(\e^{\i sh_D}\psi_j)\d s
+\xi\int_0^\infty
\sum_{x\in\cS}\langle\phi|{ G}_0^R(t,s)|x\rangle\tau_K^s(a_xV_x)\d s\\
&\sim
\int_0^\infty\sum_{x\in\cS}\langle\phi|{ G}_0^R(t,s)|x\rangle\cT_x(s)\d s
\end{align*}
where we have set
$$
\cT_x(s):=a(\e^{\i sh_D}h_T\delta_x)+\xi\tau_K^s(a_xV_x).
$$
Since ${ G}_0^R(t,s)^\ast={ G}_0^A(s,t)$, we can write
$$
\langle\phi|G^<(t,t')|\phi'\rangle
=\sum_{x,x'\in\cS}\int_0^\infty\d s\int_0^\infty\d s'
\langle\phi|{ G}_0^R(t,s)|x\rangle S^<_{xx'}(s,s')
\langle x'|{ G}_0^A(s',t')|\phi'\rangle
$$
with
$$
S^<_{xx'}(s,s')=\i\left\langle\cT_{x'}^\ast(s')\cT_x(s)\right\rangle.
$$
Note that $S^<(s,s')$, as an operator on $\fh_\cS$, is real analytic
and extends to an entire
function of $\xi$. The first terms of its Taylor expansion around $\xi=0$
$$
S^<(s,s')=\sum_{n=0}^\infty\xi^nS^{<(n)}(s,s'),
$$
are easily generated by iterating the Duhamel formula
$$
\cT_x(s)=a(\e^{\i sh_D}h_T\delta_x)
+\xi\tau_H^s(a_xV_x)
+\xi^2\int_0^s\i[\tau_H^r(W),\tau_H^s(a_xV_x)]
\d r+\mathcal{O}(\xi^3).
$$

\section{Proof of Proposition~\ref{Dyson}}\label{daison}
\subsection{Functional setup}
In this section we introduce a few function spaces which provide a convenient 
framework for the study of various GFs.

We identify $\cH_\pm=L^2(\rr_\pm,\d s)\otimes\fh$ with subspaces of
\[
\cH=L^2(\rr,\d s)\otimes\fh=\cH_-\oplus\cH_+.
\]
The Fourier transform
$
\hat{\varphi}(\omega)=\int_\rr\varphi (s)\e^{-\i\omega s}\d s$ 
maps $\cH$ to $L^2(\rr,\d\omega)\otimes\fh$ and
$\cH_\pm$ to $\fH^2(\cc_\mp)\otimes\fh$, the Hardy space of $\fh$-valued analytic 
functions on the half-plane $\cc_\mp=\{z\in\cc\,|\,\mp\Im(z)>0\}$ with the norm
\[
\|\varphi\|^2=\int_{\rr_\pm}\|\varphi(s)\|^2\d s
=\int_\rr\|\hat{\varphi}(E\mp\i0)\|^2 \frac{\d E}{2\pi}
=\sup_{\eta>0}\int_\rr\|\hat{\varphi}(E\mp\i\eta)\|^2\frac{\d E}{2\pi}.
\]
The Sobolev space $\cH^1=H^1(\rr)\otimes\fh$ (see~\eqref{iuni-1})  is a subset of $ C(\rr;\fh)$, the Banach space of continuous $\fh$-valued 
functions on $\rr$ equipped with the sup-norm. In fact elements of $\cH^1$ are 
uniformly (H{\"o}lder) continuous
\[
\|\varphi(s)-\varphi(s')\|\le|s-s'|^{1/2}\|\varphi\|_{\cH^1},
\]
\[
\sup_{s\in\rr}\|\varphi(s)\|\le\int_\rr
\|\hat{\varphi}(\omega)\|\frac{\d\omega}{2\pi}\le 2^{-1/2}\|\varphi \|_{\cH^1}. 
\]
We denote by $\cH_0^1$ the closed subspace of $\cH^1$ consisting of functions 
$\varphi:\rr\to\fh$ such that $\varphi(0)=0$.

We also introduce $\cH_\loc=L^2_\loc(\rr,\d s)\otimes\fh$, the Fr{\'e}chet space 
of all locally square integrable functions $\varphi:\rr\to\fh$ with the
seminorms
\[
\|\varphi\|_T^2=\int_{-T}^T\|\varphi(s)\|^2\d s, 
\]
and we set
\[
\cH_{\loc\pm}=\{\varphi\in\cH_\loc\,|\,\supp\varphi\subset\rr_\pm\}.
\]
We consider the subspaces
\[
\cH_\loc^1=\{\varphi\in\cH_\loc\,|\,
\chi\varphi\in\cH^1\text{ for all }\chi\in C_0^{\infty}(\rr)\},
\]
and
\[
\cH_{0,\loc}^1=\{\varphi\in\cH_\loc\,|\,
\chi\varphi\in\cH_0^1\text{ for all }\chi\in C_0^{\infty}(\rr)\},
\]
\[
\cH_{0,\loc\pm}^1=\cH_{0,\loc}^1\cap\cH_{\loc\pm}. 
\]
We shall say that a linear operator $M:\cH_{\loc\pm}\to\cH_{\loc\pm}$ is 
non-negative, and write $M\ge0$, whenever
\[
\chi_K M\chi_K\ge 0,
\]
holds as an operator on $\cH$ for all compact interval $K\subset\rr_\pm$, where 
$\chi_K$ denotes the operator of multiplication with the characteristic function 
of $K$.

\subsection{Volterra operators}

Let $\Delta=\{(s, s')\in\rr^2|0\le s'\le s <\infty\}$ and 
$\Delta\ni(s,s')\mapsto B (s,s')\in\cB(\fh)$ a continuous function such that
\[
\|B(s,s')\|\le b \e^{\gamma(s-s')}
\]
for some constants $b>0$ and $\gamma\in\rr$. The Volterra operator
with kernel $B$ is the map
\[
(V\varphi)(s)=\varphi(s)-\int_0^s B(s,s')\varphi(s')\d s',
\]
on $\cH_{\loc+}$. We shall denote it by $V=I-B$. One easily
checks that (by estimating the Hilbert-Schmidt norm)
\[
\|(I-V)\varphi\|_T
\le\frac{b}{2|\gamma|}(\e^{2\gamma T}-1- 2\gamma T)^{1/2}\|\varphi\|_T,
\]
so that $V: \cH_{\loc+}\to\cH_{\loc+}$ is continuous. Moreover, the set of 
Volterra operators on $\cH_{\loc+}$ form a group. The inverse of $V=I-B$ is the
Volterra operator $V^{-1}=I+R$ where
\[
R(s,s')=B(s,s')+
\sum_{n=2}^\infty\int_{s'\le s_1\le\cdots\le s_{n - 1}\le s}
B(s,s_{n-1})\cdots B(s_2,s_1) B(s_1, s')\d s_1\cdots\d s_{n-1},
\]
is such that
\[
\|R (s,s')\|\le b\e^{(\gamma+b)(s-s')}.
\]
In particular $\Ran\,V=\cH_{\loc+}$ and $\Ker\,V=\{0\}$.
Volterra operators on $\cH_{\loc-}$ are defined in a similar way.

\subsection{More on the non-interacting advanced/retarded Green's functions}

The operators $G_0^\pm(z):\cH\to\cH$ defined in~\eqref{aprili2} satisfy 
$G_0^\pm(z)^\ast=G_0^\mp(\bar{z})$ 
and
\[
G_0^\pm(z)-G_0^\pm(z)^\ast=(z-\bar{z})G_0^\pm(z)^\ast G_0^\pm(z)
\]
so that, in particular,
\[
\pm\Im G_0^\pm(z)=\pm\Im(z)G_0^\pm(z)^\ast G_0^\pm(z)> 0.
\]
In the Fourier representation $G_0^\pm(z)$
acts as multiplication with $(-\omega-h-z)^{-1}$ which is bounded on
$\fH^2(\cc_\pm)\otimes\fh$. It follows that
\[
G_0^\pm(z)\cH_\mp\subset\cH_\mp. 
\]
In fact the causality/anti-causality relations
\[
\supp\varphi\subset[T,\infty[
\Longrightarrow\supp G_0^-(z)\varphi\subset[T,\infty [, 
\]
\[
\supp\varphi\subset]-\infty,T]
\Longrightarrow\supp G_0^+(z)\varphi\subset]-\infty,T],
\]
hold for all $T>0$ and justify the name advanced/retarded Green's function
given to the integral kernel of $G_0^\mp$. Since $\Ran\,G_0^\pm(z)=\cH^1$ for 
$z\in\cc_\pm$, if
\[
\langle\psi|G_0^\pm(z)\varphi\rangle=0,
\]
for all $\varphi\in\cH_\mp$, then
\[
\langle G_0^\mp(\bar{z})\psi|\varphi\rangle=\langle\psi|G_0^\pm(z)\varphi\rangle 
= 0
\]
and it follows that $G_0^\mp(\bar{z})\psi\in\cH_\pm\cap\cH^1$. Since 
$(\Omega-\bar{z})G_0^\mp(\bar{z})\psi=\psi$ the locality of $\Omega$ implies that 
$\psi\in\cH_\pm$. Thus $(G_0^\pm(z)\cH_\mp)^\perp\subset\cH_\pm$ and hence
$\cH_\mp=\cH_\pm^\perp\subset(G_0^\pm(z)\cH_\mp)^{\perp\perp}$. This means that 
$G_0^\pm(z)\cH_\mp$ is dense in $\cH_\mp$. In fact one has 
$G_0^\pm(z)\cH_\mp=\cH_0^1\cap\cH_\mp$.

We also observe that the boundary values $G^\pm_0(E\pm\i0)$ are well
defined as maps from $\cH_{\loc\mp}$ to $\cH_{\loc\mp}$ with the estimate
\[
\|G_0^\pm(E\pm\i0)\varphi\|_T\le { T}\|\varphi\|_T,
\]
for all $T>0$ and $\varphi\in\cH_{\loc\mp}$. In fact
$G_0^\pm(E\pm\i0):\cH_{\loc\mp}\to\cH^1_{0,\loc\mp}$.

As an application of the above results, let us consider the inhomogeneous
Schr{\"o}dinger equation
\[
\i\partial_s\varphi (s)=(h+z)\varphi(s)+\psi(s),\qquad
\varphi(0)=f\in\fh,
\]
for $\Im z\le0$ and $\psi\in\cH_{\loc+}$. By
Duhamel's formula, the solution for $s\ge0$ is given by
\[
\varphi(s)=\e^{-\i s(h+z)}f+(G_0^-(z)\psi)(s),
\]
and thus belongs to $\cH_{\loc+}$.

\subsection{More on the interacting advanced/retarded Green's functions}

Starting from~\eqref{intGAR} we can also associate integral operators
on $\cH$ to the interacting advanced/retarded GFs. We define $G^\pm(z)$ by
$$
\langle\psi|G^\pm(z)\varphi\rangle
=\pm\i\int_{- \infty}^\infty\d s\int_{\pm(s'-s)>0}\d s' { \e^{\i (s'-s)z}}
\langle\{\tau_K^s(a(\psi(s))),
\tau_K^{s'}(a^\ast(\varphi(s')))\}\rangle,
$$
for $\pm\Im z>0$, so that
\begin{align*}
|\langle\psi|G^\pm( z)\varphi\rangle|
&\le 2\int_{- \infty}^\infty\d s\int_0^\infty\d r
\|\psi(s)\|\|\varphi(s\pm r)\|\e^{-r{ |\Im z|}}\\
&\le 2\int_0^\infty\d r\e^{-r{ |\Im z|}}\int_{-\infty}^\infty\d s\|\psi(s)\|
\|\varphi(s\pm r)\|\\
& \le 2\int_0^\infty\d r\e^{-r{ |\Im z|}} \|\psi\|\|\varphi\|\le\frac2{|\Im z|}\|\psi\|\|\varphi\|,
\end{align*}
i.e. $
\|G^\pm(z)\|\le\frac2{|\Im(z)|}$. 
Clearly, the map $z\mapsto G^\pm(z)\in\cB(\cH)$ is
analytic in $\cc_\pm$. Moreover, for $z\in\cc_+$ one has
\begin{align*}
\langle\psi|G^+(z)\varphi\rangle
&=\i\int_{-\infty}^\infty\d s\int_s^\infty\d s'\; { \e^{\i (s'-s)z}}\langle\{
\tau_K^s(a(\psi(s))),\tau_K^{s'}(a^\ast(\varphi(s')))
\}\rangle\\
&=\i\int_{-\infty}^\infty\d s'\int_{-\infty}^{s'}\d s\; 
{ \e^{\i (s'-s)z}}
\langle\{
\tau_K^s(a(\psi(s))),\tau_K^{s'}(a^\ast(\varphi(s'))) 
\}\rangle\\
&=\i\int_{-\infty}^\infty\d s\int_{-\infty}^s\d s'\; { \e^{-\i (s'-s)z}}\langle \{
\tau_K^{s'}(a(\psi(s'))),\tau_K^s(a^\ast(\varphi(s))) 
\}\rangle\\
&=\overline{\left[-\i\int_{-\infty}^\infty\d s\int_{-\infty}^s\d s'\; { \e^{\i (s'-s)\bar{z}}}\langle \{ 
\tau_K^{s'}(a^\ast(\psi(s'))),\tau_K^s(a(\varphi(s))) \}\rangle\right]}\\
&=\overline{\langle\varphi|G^-(\bar{z})\psi\rangle}=\langle G^-(\bar{z})\psi|\varphi\rangle,
\end{align*}
hence
\beq\label{aprili3}
G^\pm(z)^\ast=G^\mp(\bar{z}).
\eeq
Finally, it immediately follows from its definition that $G^\pm(z)$
satisfies the same causality/anti-causality relations than $G_0^\pm(z)$,
and in particular that
\[
G^\pm(z)\cH_\mp\subset\cH_\mp.
\]

\subsection{The reducible self-energy}

For $\varphi\in\cH^1$, we can write
$$
\langle\psi|G^\pm(z)(\Omega-z)\varphi\rangle
=\pm\i\int_{-\infty}^\infty\d s\int_0^\infty\d r 
\langle\{\tau_K^s(a(\e^{\i s\bar{z}}\psi(s))), 
\tau_K^{s\pm r}(a^\ast(\tilde{\varphi}(s\pm r)))\}\rangle
$$
where $\tilde{\varphi}(s')=\e^{\i s'z}((\Omega-z)\varphi)(s')$. Let us define
$b(f):=\i\xi[W,a (f)]$, so that $b^\ast(f)=\i\xi [W,a^\ast(f)]$ and
$$
\i[K,a^\#(f)]=\partial_s\tau_K^s(a^\#(f))|_{s=0}=a^\#(\i hf)+b^\#(f).
$$
One easily derives the relations
\begin{equation}
\tau_K^t(a^\ast(\tilde{\varphi}(t)))
=\i\partial_t\tau_K^t (a^\ast(\e^{\i tz}\varphi(t)))
-\i\tau_K^t(b^\ast(\e^{\i tz}\varphi(t)))
\label{Eq1}
\end{equation}
which leads to:
$$
\tau_K^{s\pm r}(a^\ast(\tilde{\varphi}(s\pm r)))
=\pm \i\partial_r\tau_K^{s\pm r} (a^\ast(\e^{\i (s\pm r)z}\varphi(s\pm r)))
-\i\tau_K^{s\pm r}(b^\ast(\e^{\i (s\pm r)z}\varphi(s\pm r))).
$$
Integration by parts with respect to $r$ yields:
$$
\langle\psi|G^\pm(z)(\Omega-z)\varphi\rangle
=\langle\psi|\varphi\rangle\pm\int_{- \infty}^\infty\d s\int_0^\infty\d r
\langle\{\tau_K^s(a(\e^{\i s\bar{z}}\psi(s))),
\tau_K^{s\pm r}(b^\ast(\e^{\i(s\pm r)z}\varphi(s\pm r))\} \rangle,
\label{Eq2}
$$
which we can rewrite as
\[
G^\pm(z)(\Omega-z)=I+F^\pm(z),
\]
where $F^\pm(z)$ extends to a bounded operator on $\cH$ with
\[
\|F^\pm(z)\|\le\frac{4|\xi|\|W\|}{|\Im z|}.
\]
Using this together with~\eqref{aprili3}, by duality it follows that 
$\Ran\,G^\pm(z)\subset\cH^1$. For $\psi\in\cH^1$, we further have
\[
\langle(\Omega-\bar{z})\psi|F^\pm(z)\varphi\rangle
=\pm\int_{-\infty}^\infty\d s\int_{\pm (s'-s)>0}\d s'
\langle\{\tau_K^s(a(\tilde{\psi}(s))),
\tau_K^{s'}(b^\ast(\e^{\i s'z}\varphi(s'))\}\rangle,
\]
where
\[
\tilde{\psi}(s)=\e^{\i s\bar{z}}((\Omega-\bar{z})\psi)(s).
\]
Using the adjoint of~\eqref{Eq1}:
$$
\tau_K^s(a(\tilde{\psi}(s)))=-\i\partial_s\tau_K^s(a(\e^{\i s\bar{z}}\psi(s)))
+\i\tau_K^s(b(\e^{\i s\overline{z}}\psi(s)),
\label{Eq3}
$$
interchanging the $s$ and $s'$ integrals and then  integrating by parts with 
respect to $s$, yields
\[
\langle(\Omega-\bar{z})\psi|F^\pm(z)\varphi\rangle
=\langle\psi|\widetilde{\Sigma}^\pm(z)\varphi\rangle,
\]
where
\[
\widetilde{\Sigma}^\pm(z)=v_{\rm HF}+\fS^\pm(z),
\]
with
\[
(v_{\rm HF}\varphi)(s)=v_{\rm HF}(s)\varphi(s)
\]
the Hartree-Fock energy
$$
(v_{\rm HF}(s)f)(x)=-\i\langle\tau_K^s(\{a_x,b^\ast(f)\})\rangle,\quad \langle\delta_y|v_{\rm HF}(s)\delta_x\rangle
=\xi\langle\tau_K^s(\{a_y,[W,a_x^\ast]\})\rangle,
\label{Eq4}
$$
and
$$
\langle\psi|\fS^\pm(z)|\varphi\rangle
=\pm\i\int_{- \infty}^\infty\d s\int_{\pm(s'-s)>0}\d s'\langle\{
\tau_K^s(b(\e^{\i s\bar{z}}\psi(s))),\tau_K^{s'}(b^\ast(\e^{\i s'z}\varphi(s'))) 
\}\rangle.
$$
We note that $v_{\rm HF}(s)$ is a finite rank operator on 
$\fh$ with $\Ran\,v_{\rm HF}(s)\subset\fh_\cS$, 
and an entire analytic function of $s$. In the NESS, it is independent of $s$. 
Let us show that it is self-adjoint. We have: 
$$
\langle y|v_{\rm HF}(s)|x\rangle
=\xi\langle\tau_K^s(a_y Wa_x^\ast -a_y a_x^\ast W +Wa_x^\ast a_y 
- a_x^\ast Wa_y)\rangle
$$
and 
$$
\overline{\langle y|v_{\rm HF}(s)|x\rangle}
=\xi\langle\tau_K^s(a_x Wa_y^\ast - Wa_x a_y^\ast +a_y^\ast  a_x W
- a_y^\ast Wa_x)\rangle.
$$
Using that $\{a_x,a_y^*\}=\{a_x^*,a_y\}=\delta_{xy}$ we obtain 
$\langle y|v_{\rm HF}(s)|x\rangle
=\overline{\langle x|v_{\rm HF}(s)|y\rangle}$ and we are done.
 
The operators $\fS^\pm(z)$ define functions 
$(s,s')\mapsto\fS^\pm(z|s,s')\in\cB(\fh_\cS)$ which, up to a factor of
$\theta(\pm(s'-s))$ are also entire analytic in both variables
$$
\langle g|\fS^\pm(z|s,s')|f\rangle
=\pm\i\theta(\pm(s'-s))\e^{\i(s'-s)z}\langle\{\tau_K^s(b(g)),
\tau_K^{s'}(b^\ast(f))\}\rangle.
$$
These operators are also finite rank with
\[
\Ran\,\fS^\pm(z|s,s')\subset\fh_\cS,\qquad
\fh_\cR\subset\Ker\,\fS^\pm(z|s,s'),
\]
and
\[
\langle\psi|\fS^\pm(z)|\varphi\rangle
=\int_{-\infty}^\infty\d s\int_{-\infty}^\infty\d s'\langle
\psi(s)|\fS^\pm(z|s,s')|\varphi(s')\rangle.
\]
The relation
\[
\fS^+(z)^\ast=\fS^-(\bar{z}),
\]
can be proved following the same idea as in~\eqref{aprili3}. In the NESS, $\fS^\pm(z|s,s')=\fS^\pm(z|s-s')$.

Summarizing, we have shown that $G^\pm(s)(\Omega-z)=I+F^\pm(z)$ and 
$(\Omega-z)F^\pm(z)=\widetilde{\Sigma}^\pm(z)$ which
can be rewritten as $G^\pm(z)=G_0^\pm(z)+F^\pm(z)G_0^\pm(z)$ and 
$F^\pm(z)=G_0^\pm(z)\widetilde{\Sigma}^\pm(z)$. The
operator $\widetilde{\Sigma}^\pm(z)$ is called {\sl reducible
advanced/retarded self-energy.} Thus, we have proved

\begin{lemma}\label{lemaaprili}
For $\pm\Im z>0$ the advanced/retarded interacting Green's
function satisfies the equation
\begin{equation}
G^\pm(z)=G_0^\pm(z)+G_0^\pm(z)\widetilde{\Sigma}^\pm(z)G_0^\pm(z),
\label{EqG}
\end{equation}
where $\widetilde{\Sigma}^\pm(z)$ is the reducible self-energy, an operator 
acting on $L^2(\rr,\d s)\otimes\fh_\cS$ as
\[
(\widetilde{\Sigma}^\pm(z)\varphi)(s)=v_{\rm HF}(s)\varphi(s)
+\int\fS^\pm(z|s,s')\varphi(s')\d s',
\]
with
\[
\langle y|v_{\rm HF}(s)|x\rangle
=\xi\langle\tau_K^s(\{a_y,[W,a_x^\ast]\})\rangle, 
\]
and
\[
\langle y|\fS^\pm(z|s,s')|x\rangle
=\mp\i\xi^2\theta(\pm(s'-s))\e^{\i( s'-s)z}\langle\{
\tau_K^s([W,a_y]),\tau_K^{s'}([W,a_x^\ast])\}\rangle.
\]
\end{lemma}

We observe that the boundary values of all these advanced/retarded operators
$G^\pm_0(E\pm\i0)$, $G^\pm(E\pm\i0)$ and $\tilde{\Sigma}^\pm(E\pm\i0)$ are well 
defined as operators on $\cH_{\loc\mp}$, by the same estimate as for the free 
Green's functions. It follows that relation~\eqref{EqG} remains valid 
on $\cH_{\loc\mp}$ for $z=E\pm\i 0$. In particular
$$
\langle x|G^{A/R}(s,s')|x'\rangle=\langle x|G^{A/R}_0(s,s')|x'\rangle
+\sum_{y,z\in\cS}\int\d u\int\d v\,\langle x|G^{A/R}_0(s,u)|y\rangle
\widetilde{\Sigma}^\pm_{yz}(0\pm\i0|u,v)\langle z|G^{A/R}_0(v,s')|x'\rangle.
$$

\begin{remark}
The self-energy $\widetilde{\Sigma}^\pm(z)$ has an important property
related to dissipation. As already noticed, the Hartree-Fock part is
self-adjoint, thus, for $z=E-\i\eta$, $\eta>0$ and $\varphi\in\cH_+$, one has
\[
(\tilde{\Sigma}^-(z)\varphi)(s)=v_{\rm HF}(s)\varphi(s)
+\int_0^s\fS^-(z|s,s')\varphi(s')\d s',
\]
and hence, with $B_s=\tau_K^s(b(\e^{\i sE}\varphi(s)))$ and invoking
Lemma~\ref{LemmaPositive},
\begin{align}
\Im\int_0^T\langle\varphi(s)|(\tilde{\Sigma}^-(z)\varphi)(s)\rangle\d s 
&=\Im\int_0^T\d s\int_0^s\d s'\langle\varphi(s)|\fS^-(z|s,s')|\varphi(s')\rangle \nonumber\\
&=-\Re\int_0^T\d s\int_0^s\d s'\langle\{\tau_K^s(b(\e^{\i s\bar{z}}\varphi(s))),
\tau_K^{s'}(b^\ast(\e^{\i s'z}\varphi(s')))\}\rangle\nonumber \\
&=-\Re\int_0^T\d s\int_0^s\d s'\e^{-\eta(s-s')}\langle\{B_s,B_{s'}^\ast\}\rangle 
\le0.\label{aprili4}
\end{align}
\end{remark}

\subsection{The irreducible advanced/retarded self-energy}
Considering $G_0^\pm$ and $G^\pm$ as operators on $\cH_{\loc\mp}$, the relation
\begin{equation}
G^\pm(z)=G_0^\pm(z)+G_0^\pm(z)\widetilde{\Sigma}^\pm(z)G_0^\pm(z),
\label{Eq5}
\end{equation}
can be rewritten as
\[
G^\pm(z)=\left (I+G_0^\pm(z)\widetilde{\Sigma}^\pm(z)\right )G_0^\pm(z),
\]
and since $I+G_0^\pm(z)\widetilde{\Sigma}^\pm(z)$ is a Volterra
operator on $\cH_{\loc\mp}$, the last relation leads to
\[
G_0^\pm(z)=\left (I+G_0^\pm(z)\widetilde{\Sigma}^\pm(z)\right )^{-1}G^\pm(z),
\]
which, inserted into~\eqref{Eq5}, yields
\begin{equation}
G^\pm(z)=G_0^\pm(z)+G_0^\pm(z)\Sigma^\pm(z)G^\pm(z),
\label{Eq6}
\end{equation}
with
\begin{equation}
\Sigma^\pm(z)=\widetilde{\Sigma}^\pm(z)
\left (I+G_0^\pm(z)\widetilde{\Sigma}^\pm(z)\right )^{-1},
\label{Eq8}
\end{equation}
a Volterra kernel. The operator $\Sigma^\pm(z)$ is the {\sl irreducible 
self-energy} announced in~\eqref{aprili1}. This ends the proof of 
Proposition~\ref{Dyson}.

\subsection{Some dissipative properties of the irreducible self-energy}
Since $I-\Sigma^\pm(z)G^\pm(z)$ is a Volterra
operator, \eqref{Eq6} implies that $G^\pm(z)\cH_\mp$ is dense in
$\cH_\mp$. Multiplying~\eqref{Eq6} on the left by $\Omega-z$
further gives
\[
\left (\Omega-\Sigma^\pm(z)-z\right )G^\pm(z)=I,
\]
so that
\[
G^\pm(z)=(\Omega-\Sigma^\pm(z)-z)^{-1}. 
\]
Reasoning like in~\eqref{aprili4} we obtain $
\Im(G^-(z))\le 0$. Using this in the `resolvent identity'
\[
G^-(z)-G^-(z)^\ast=G^-(z)^\ast(z-\bar{z}+\Sigma^-(z)-\Sigma^-(z)^\ast)G^-(z) 
\]
one deduces
\begin{align}\label{iuni-2}
\Im(z+\Sigma^-(z))\le 0 
\end{align}
and in particular, in the limit $\Im z\to 0$,
\[
\Im(\Sigma^-(E-\i0))\le 0.
\]
Let us consider the Schr{\"o}dinger equation
\[
\i\partial_s\varphi(s)=(h+z)\varphi(s)+(\Sigma^-(z)\varphi)(s)
\]
with the initial condition $\varphi(0)$ for $\Im z\le0$.
The  dissipative property~\eqref{iuni-2} ensures that the
solution satisfies $\|\varphi(s)\|\le\|\varphi(0)\|$ for $s\ge0$:
\[
\frac12\partial_s\| \varphi(s)\|^2=\Im\langle\varphi(s)|
\i\partial_s\varphi(s)\rangle=\Im\langle\varphi(s)|
(z+\Sigma^-(z)\varphi)(s)\rangle
\]
so
\[
\frac12(\|\varphi(T)\|^2-\|\varphi(0)\|^2)
=\Im\int_0^T\d s\int_0^s\d s'\langle\varphi(s)| \Sigma^-(z|s,s')|\varphi(s')\rangle
+\Im z\int_0^T\|\varphi(s)\|^2\d s\le0.
\]
It follows that the equation has a contractive propagator
$\varphi(s)=U(s,s')\varphi(s')$. By Duhamel's formula, the initial value problem 
is equivalent to the integral equation
\begin{align*}
\varphi(s)&=\e^{-\i s(h+z)}\varphi(0)+(G_0^-(z)\Sigma^-(z)\varphi)(s)\\
&=\e^{-\i s(h+z)}\varphi(0)-\i\int_0^s\d r\e^{-\i(s-r)(h+z)}
\int_0^r\d s'\Sigma^-(r,s')\varphi(s').
\end{align*}
Since $G_0^-(z)\Sigma^-(z)$ is a Volterra kernel, this equation is solved
by setting
\[
\varphi(s)=((I-G_0^-(z)\Sigma^-(z))^{-1}\psi)(s),
\]
with
\[
\psi(s)=\theta(s)\e^{-\i s(h+z)}\varphi(0)=\i G_0^-(z)\delta_0\otimes\varphi,
\]
where $\delta_0\otimes\varphi\in\cH^{-1}$, $\delta_0$ denoting
the Dirac mass at $s=0$. By~\eqref{Eq8}
\begin{align*}
I-G_0^-(z)\Sigma^-(z)&=I-G_0^-(z)\widetilde{\Sigma}^-(z)
(I+G_0^-(z)\widetilde{\Sigma}^-(z))^{-1}\\
&=(I+G_0^-(z)\widetilde{\Sigma}^-(z))^{-1}\\
\varphi(s)&=(I+G_0^-(z)\widetilde{\Sigma}^-(z))\psi\\
&=\i(I+G_0^-(z)\widetilde{\Sigma}^-(z))G_0^-(z)\delta_0\otimes \varphi\\
&=\i G^-(z)\delta_0\otimes \varphi.
\end{align*}
Thus we have the following formula for the interacting Green's function
\[ 
(G^-(z)\varphi)(s)=-\i\int_{-\infty}^s U (s,s')\varphi(s')\d s' .
\]

\section{Concluding remarks and open problems}\label{concluzii}
We established the first systematic mathematical approach to the non-equilibrium
Green's function formalism for interacting transport in open systems. Rather
than introducing the textbook Keldysh contours and contour-ordering operators we
follow a three-step bottom-up strategy only using  real-time GFs (i.e., retarded,
advanced and lesser):
\begin{enumerate}
\item We relate the time-dependent current to a fully interacting lesser
GF $\langle \phi_j|G^<(t,t)|\psi_j\rangle$ associated with a pair of states from a 
given lead $j$ and the sample.

\item Using the KMS condition and Duhamel identities we show that 
$\langle \phi_j|G^<(t,t)|\psi_j\rangle$ obeys a Langreth-type identity which 
immediately implies the JMW formula.

\item We derive the Keldysh equation for the lesser GF restricted to the sample 
in terms of a lesser interaction self-energy for which we provide explicit 
expressions. 
\end{enumerate}

Last but not least we rely on the theory of Volterra operators to rigorously
define the irreducible retarded self-energy via its Dyson equation. This is a
mandatory back-up for practical diagrammatic recipes (e.g., Hartree-Fock,
self-consistent Born approximation, $GW$). For completeness and comparison we
recall here that in the standard top-down way to the JMW formula one rather
takes for granted a Dyson equation for contour-ordered GFs and then uses the
formal Langreth rules to come back to real-time quantities.

Let us also point out some open problems related to the NEGF formalism for 
interacting systems:

$\bullet$ {\it Initial correlations.} We recall that the Keldysh
identity~\eqref{iulie1401} requires that the initial state of the sample is the
vacuum. From a physical point of view, this means that there are no initial
correlations due to the Coulomb interaction between particles before the
coupling to the leads is established. If the initial state can have particles in
the sample before the coupling, the Keldysh equation  acquires a more
complicated form~\cite{HJ, SvL}.

$\bullet$ {\it The partition-free setting.} This alternative transport scenario
has been put forward a long time ago by Cini~\cite{Ci} and goes as follows: i)
the initial state describes a coupled but unbiased system where all the chemical
potentials $\mu_j$ are equal, i.e., the initial state $\langle
\cdot\rangle^{\beta,\mu}$ is the thermodinamic limit of the grand canonical
Gibbs state associated with the restriction of $K$ to finitely extended
reservoirs; ii) at some instant $t_0$  one adds a potential bias  on the leads
and a current is established. Later on, Stefanucci and Almbladh~\cite{SA}
adapted the Keldysh formalism for the partition-free scenario. They found in
particular that the Keldysh equation for the lesser interaction self-energy  is
far more complicated. In fact, they argue that in this case the Keldysh equation
should only be used to describe the long-time response of the interacting
system.  They also derive a  generalized JMW formula (see Eq. (16.7)
in~\cite{SvL}). Although we~\cite{CMP1,CMP2} were able to establish the existence
of a NESS {\it in the fully resonant case} for the free-partition setting as
well, a `contourless' derivation of a JMW-type formula and its corresponding
Langreth-like identities is still missing.

\appendix

\section{ A positivity lemma}

\begin{lemma}\label{LemmaPositive}
Let $[0,T]\ni s\mapsto A_s\in\cB(\fh)$ be a continuous map 
such that $\sup_{s\in[0,T]}\|A_s\|< \infty$. Then for any $\eta\ge 0$ one has
\[
\Re\int_0^T\d s\int_0^s\d s'\e^{-\eta (s - s')}A_s^\ast A_{s'}\ge0.
\]
\end{lemma}
{\noindent}{\bf Proof.} It suffices to show that for any $f\in\fh$
\[
\Re\int_0^T\d s\int_0^s\d s'\e^{-\eta(s-s')}\langle f_s|f_{s'}\rangle\ge0,
\]
where $f_s=A_sf$. Extending $s\mapsto f_s$ by zero on $\rr\setminus[0,T]$ 
yields a strongly measurable $\fh$-valued function which belongs to 
$L^1(\rr;\fh)$. Set
\[
f_{s,\epsilon}=\int_0^T\e^{-(s-s')^2/2\epsilon}f_{s'}
\frac{\d s'}{\sqrt{2\pi\epsilon}}, 
\]
so that $f_{s,\epsilon}\to f_s$ in $L^1(\rr;\fh)$
as $\epsilon\downarrow0$. It follows from the dominated convergence theorem
that
\[
\Re\int_0^T\d s\int_0^s\d s'\e^{-\eta(s-s')}\langle f_s|f_{s'}\rangle
=\lim_{\epsilon\downarrow 0}\Re\int_{-\infty}^{\infty}\d s\int_{-\infty}^s\d s' 
\e^{-\eta(s-s')}\langle f_{s,\epsilon}|f_{s',\epsilon}\rangle . 
\]
The Fourier transform of $f_{s,\epsilon}$ is given by 
$\e^{-\epsilon\omega^2/2}\widehat{f}_\omega$ with
\[
\widehat{f}_\omega=\int_0^T\e^{-\i\omega s}f_s\d s.
\]
It follows that $(\nu,\omega)\mapsto\langle\widehat{f}_{\nu,\epsilon}
|\widehat{f}_{\omega,\epsilon}\rangle$ is a Schwartz function and an explicit
calculation yields
$$
\int_{-\infty}^{\infty}\d s\int_{-\infty}^s\d s'\e^{-\eta(s-s')}
\langle f_{s,\epsilon}|f_{s',\epsilon}\rangle
=\int_{-\infty}^{\infty}\frac{\d\omega}{2\pi}\frac{1}{\eta-i\omega}
\|\widehat{f}_{\omega,\epsilon} \|^2, 
$$
so that
\[
\Re\int_{-\infty}^{\infty}\d s\int_{-\infty}^s\d s'
\e^{-\eta(s-s')}\langle f_{s,\epsilon}|f_{s',\epsilon}\rangle
=\int_{-\infty}^{\infty}\frac{\d\omega}{2\pi}\frac{\eta}{\eta^2+\omega^2}
\|\widehat{f}_{\omega,\epsilon}\|^2 \ge 0.
\]

\end{document}